\documentclass[aps,prl,reprint,superscriptaddress,nofootinbib,floatfix,showkeys,longbibliography]{revtex4-2}
\pdfoutput=1

\usepackage[utf8]{inputenc}
\usepackage[T1]{fontenc}

\usepackage{amsmath}
\usepackage{amssymb}
\usepackage{graphicx}
\usepackage{dcolumn}
\usepackage{xcolor}
\usepackage{float}
\usepackage{bm}
\usepackage{comment}
\usepackage{tabularx}
\usepackage{subfigure}
\usepackage{microtype}

\usepackage[italicdiff]{physics}
\usepackage[colorlinks=True,allcolors=blue]{hyperref}

\allowdisplaybreaks

\newcommand{\prob}{\operatorname{Prob}}
\newcommand{\sgn}{\operatorname{sgn}}

\newcommand{\imag}{\operatorname{Im}}
\newcommand{\sinc}{\operatorname{sinc}}

\newcommand\MAxGeneva{Section of Mathematics, University of Geneva, Rue du Conseil-G\'en\'eral 7-9, 1205 Geneva, Switzerland}

\newcommand\QPC{Quantum Physics Corner Ltd, 20–22 Wenlock Road, London N1 7GU, United Kingdom}

\newcommand\UGxIFTiA{Institute of Theoretical Physics and Astrophysics,
Faculty of Mathematics, Physics and Informatics,
University of Gda{\'n}sk, 
80-308 Gda{\'n}sk, Poland}

\newcommand\XMUMxPH{School of Mathematics and Physics, 
Xiamen University Malaysia, 
43900 Sepang, Malaysia}
\begin{document}

%%%%%%%%%%%%%%%%%%%%%%%%%%%%%%%%%%%%
\title{General quantum backflow in realistic wave packets}

%%%%%%%%%%%%%%%%%%%%%%%%%%%%%%%%%%%%

\author{Tomasz Paterek}
\affiliation{\XMUMxPH}
\affiliation{\UGxIFTiA}

\author{Arseni Goussev}
\affiliation{\MAxGeneva}
\affiliation{\QPC}

%%%%%%%%%%%%%%%%%%%%%%%%%%%%%%%
\begin{abstract}
	Quantum backflow is a counterintuitive phenomenon in which the probability density of a quantum particle propagates opposite to its momentum. Experimental observation of backflow has remained elusive due to two main challenges: (i) the effect is intrinsically small, with less than 4\% of the probability able to flow backward, and (ii) it requires wave packets with a well-defined momentum direction, which are difficult both to prepare and to verify under realistic, noisy conditions. Here, we overcome these challenges by introducing a general formulation of quantum backflow applicable to arbitrary momentum distributions. The framework recovers the standard backflow limit for unidirectional states and identifies general backflow as probability flow exceeding that predicted by the particle’s momentum distribution alone. We show that this excess can reach nearly 13\%, surpassing the standard backflow bound by more than a factor of three. Furthermore, we extend the framework to the closely related phenomenon of quantum reentry, provide explicit examples of quantum states exhibiting large general backflow and reentry, and discuss the foundational implications of these nonclassical effects. Our results open a pathway toward the experimental observation of quantum backflow in realistic settings.
\end{abstract}
%%%%%%%%%%%%%%%%%%%%%%%%%%%%%%%

%%%%%%%%%%%%%%%%%%%%%%%%%%%%%%%
\maketitle
%%%%%%%%%%%%%%%%%%%%%%%%%%%%%%%

\emph{Introduction.} Interference of matter waves---evidenced by the appearance of characteristic density fringes obtained through repeated position measurements of a particle at a given time---has been a cornerstone of quantum mechanics since its early days. Only relatively recently, however, has it been recognized that performing position measurements at multiple distinct times can reveal nonclassical aspects of probability transport even for a free nonrelativistic particle. Perhaps the most well-known example is quantum backflow~\cite{Allcock1969, Kijowski1974, Wer88Wigner, Bracken1994,Bracken2021}, but other notable effects include quantum reentry~\cite{Goussev2019}, Tsirelson’s precession~\cite{Tsirelson2006, Zaw2022, Zaw2025prl, Zaw2025, Vaartjes2025,Garg2025_arXiv}, and quantum advantage in various transportation tasks~\cite{Hof17Quantum, Hof21How, Ono2023,  Trillo2023}. These effects appeal to the very core of quantum mechanics, much like interference and diffraction. Yet even quantum backflow, which manifests as a classically impossible transfer of probability against the particle’s momentum, has not yet been experimentally observed.\footnote{There has been significant experimental progress in studying optical analogs of quantum backflow, where the instantaneous local momentum of a light field is shown to point opposite to the local momentum of its constituent modes. Such effects have been demonstrated in linear motion using classical optical beams~\cite{Eliezer2020, Daniel2022}, and in rotational motion using both classical light~\cite{Ghosh2023} and single photons~\cite{ZHD+25Observation}. However, because these experiments are effectively time-independent, they do not capture the probability dynamics that fundamentally define quantum backflow.}

Two main obstacles have hindered the experimental observation of quantum backflow. The first arises from the intrinsic weakness of the effect. Bracken and Melloy~\cite{Bracken1994} established a limit to the probability that can flow against the particle's momentum---the Bracken–Melloy constant $c_{\text{BM}}$---which they estimated to be about $4\%$. This estimate has been progressively refined~\cite{EFV05Quantum, PGKW06new}, with the most accurate value being $c_{\text{BM}} = 0.0384506$~\cite{FK--Repeated}. (Exact but not optimal analytical bounds on $c_{\text{BM}}$ are reported in Refs.~\cite{Trillo2023,Zaw2025}.) However, the state saturating this limit is highly singular: it possesses a discontinuous momentum distribution, infinite energy, and diverging probability current, making it physically unrealistic. More regular wave functions~\cite{Bracken1994, EFV05Quantum, Yearsley2012, HGL+13Quantum, Palmero2013, Miller2021, BG21experiment, Chr24Design} yield substantially smaller backflow values, typically below $0.016$, rendering the effect challenging to detect.

The second obstacle is the strict requirement of a fixed momentum direction. Such states are difficult to engineer and verify because realistic, noisy measurements inevitably produce spurious outcomes that can indicate an incorrect momentum direction even for properly prepared initial states. Moreover, in some systems---such as ballistic electrons in mesoscopic conductors---preparing a strictly unidirectional wave packet is fundamentally impossible~\cite{Goussev2025}. 
While Bracken has generalized the standard backflow formulation to systems with any momentum cutoff~\cite{Bracken2021}, these considerations motivate the development of a broader framework capable of describing quantum backflow for arbitrary momentum distributions.

Here, we develop such a general formulation, applicable to both quantum backflow and the closely related phenomenon of quantum reentry. The latter describes the nonzero probability that a free quantum particle reenters a spatial region it has previously left---a behavior forbidden to classical free particles, whose motion is unidirectional in the absence of external forces. Backflow emerges as a special case of reentry~\cite{Gou20Probability}. We show that, when extended to arbitrary momentum distributions, quantum reentry can be tested by examining only a finite region of the experimental interferogram, making this approach particularly suitable for experimental implementation.

Our approach is founded on the observation that quantum interference enables a particle to reach spatial regions inaccessible to classical particles. We express the classical   limits on probability transport as two inequalities---one for backflow and one for reentry---that make no assumptions about momentum direction or spatial localization. Quantum mechanics can violate these inequalities, with up to $0.1281$ of the particle's probability flowing opposite to the classical expectation. The optimal state yielding this maximal violation differs from the standard backflow-maximizing state. We also investigate general backflow in experimentally relevant settings using superpositions of Gaussian wave packets. While Gaussian states have been considered in prior studies of standard backflow~\cite{Yearsley2012}, only the present work provides a rigorous treatment that accounts for their infinite momentum support.

%%%%%%%%%%%%%%%%%%%%%%%%%%%%%%%%%%%%%%%%%%%

\emph{Classical limits.} We begin by characterizing the absence of backflow in classical systems. Consider a free classical particle of mass $m$ moving along the $x$ axis, with an unrestricted phase-space distribution $f(x,p,t) = f(x - pt/m, p, 0)$. The central quantity in backflow-like problems is the probability of finding the particle to the left of the origin $x = 0$ at time $t$, denoted as $P_-(t) = \prob(x < 0, t)$. This probability can change between times $t_1$ and $t_2 = t_1 + T$ in two distinct ways: (i) particles with positive momentum ($p$ > 0) initially located in the interval $x \in (-pT/m, 0)$ will leave the region, while (ii) particles with negative momentum ($p < 0$) initially positioned in $x \in (0, -pT/m)$ will enter it. Since in general backflow we are interested in the increment of this probability, we upper-bound the probability gain by the second contribution only:
\begin{equation}
	\hspace{-0.3cm}  P_-(t_2) - P_-(t_1) \le \int_{-\infty}^0 \!\! dp \int_0^{- \frac{pT}{m}} \!\! dx \, f(x,p,t_1) \le \tilde P_- \,.
	\label{EQ_B}
\end{equation}
Here, $\tilde{P}_- = \prob(p < 0)$ is the probability of negative momentum, which remains time independent under free evolution. %The integrals in Eq.~(\ref{EQ_B}) represent contribution (ii), and 
The second inequality follows by extending the $x$-integration limits to the entire real line and using the non-negativity of the probability density $f$.

Inequality~(\ref{EQ_B}) is expressed solely in terms of measurable position and momentum marginals, making it directly comparable to quantum results. For $\tilde{P}_- = 0$, it reduces to the familiar statement that the probability of finding a right-moving particle to the left of the origin cannot increase with time. In its general form, however, the inequality makes no assumptions about the marginals and simply bounds the position-probability gain by the probability of negative momentum. Any violation of this bound---shown below to be possible for quantum wave packets---therefore signals a nonclassical transport effect where the particle moves backward more than is permitted by its momentum distribution.

The momentum probability in (\ref{EQ_B}) can be traded for a position probability to produce a generalization of the reentry problem~\cite{Goussev2019}. Consider a finite interval along the $x$ axis, $r = (-s, 0)$ for concreteness, and define $P_r(t) = \prob(x \in r, t)$ as the probability for the particle to be inside $r$ at time $t$. The difference $P_r(t_2) - P_r(t_1)$ quantifies the fraction of particles that have entered $r$ between times $t_1$ and $t_2 > t_1$. Since classical particles can only enter from outside the region, this increase cannot exceed the total fraction of particles that were outside $r$ at any earlier time $t_0 < t_1$, namely $1 - P_r(t_0)$. This leads to the following inequality characterizing the impossibility of reentry for classical particles:
\begin{equation}
	P_r(t_2) - P_r(t_1) \le 1 - P_r(t_0) \,,
	\label{EQ_RE}
\end{equation}
which holds for any $t_0 < t_1 < t_2$.\footnote{See Supplemental Material for an alternative derivation of Eqs.~\eqref{EQ_B} and \eqref{EQ_RE}.} For $P_r(t_0) = 1$ and $s \to \infty$, it reduces to the classical reentry condition discussed in Refs.~\cite{Goussev2019, Gou20Probability}.

Compared to~\eqref{EQ_B} the reentry inequality offers two notable advantages. First, it depends solely on position measurements, allowing it to be tested experimentally by comparing interference patterns at three distinct times. Second, only a finite region of the interferogram, corresponding to $r$, needs to be analyzed, provided the total flux of particles is known.

\emph{Simple example.} We now demonstrate that the classical bounds expressed in inequalities~\eqref{EQ_B} and~\eqref{EQ_RE} can be violated by quantum particles. Consider a wave function consisting of a superposition of two Gaussian wave packets, $\Psi(x,t) = \sum_{n=1,2} c_n g_n(x,t)$, directly accessible in optical systems and Bose-Einstein condensates (BECs). This type of superposition was previously analyzed in Ref.~\cite{Yearsley2012} for real coefficients $c_n$ and Gaussians $g_n(x,t)$ with identical widths, yielding quantum backflow of $0.0061$. Although strictly nonzero, the corresponding probability of negative momentum was found to be small, $\tilde{P}_- \approx 10^{-10}$, and was therefore neglected.

To systematically study the effect of negative momentum, we define
\begin{eqnarray}
	\Delta_{\mathrm{QB}} & = & P_-(t_2) - P_-(t_1) - \tilde P_- \,, \label{EQ_DELTA_QB} \\
	\Delta_{\mathrm{RE}} & = & P_-(t_2) - P_-(t_1) + P_-(t_0) - 1 \,, \label{EQ_DELTA_RE}
\end{eqnarray}
which, according to Eqs.~\eqref{EQ_B} and~\eqref{EQ_RE}, cannot exceed zero in classical mechanics. Details of maximizing these quantities over the set of superposed Gaussian states are provided in the Supplemental Material (SM). Using the same assumptions as in Ref.~\cite{Yearsley2012}, we find that the generalized quantum backflow reaches $\Delta_{\mathrm{QB}} = 0.0106$. Relaxing the assumptions on the Gaussian superposition increases the maximum backflow to $\Delta_{\mathrm{QB}} = 0.0120$. The same enhancements are observed for quantum reentry, $\Delta_{\mathrm{RE}}$. While these values remain modest, they clearly demonstrate that incorporating negative-momentum components allows substantially larger violations of classical probability bounds.

\emph{Large backflow example.} The previous illustration showed that general quantum backflow can exceed the magnitude observed in the standard scenario. Remarkably, general backflow can even significantly surpass the Bracken–Melloy constant, $c_{\text{BM}} = 0.0384506$. Inspired by Ref.~\cite{HGL+13Quantum}, we construct an explicit example using the momentum-space wave function $\tilde{\psi}(p) = \Theta(p) a_1 (b_1 - p) e^{-c_1^2 p^2} + \Theta(-p) a_2 (b_2 - p) e^{-c_2^2 p^2}$, where $a$’s, $b$’s, and $c$'s are constants, and $\Theta(\cdot)$ is the Heaviside step function. Such superpositions are feasible in BEC platforms. With a suitable choice of parameters, this state achieves a general backflow of $\Delta_{\mathrm{QB}} = 0.0624$. There also exists a corresponding state exhibiting the same value of $\Delta_{\mathrm{RE}}$. Detailed derivations and parameter choices are provided in the SM.

\emph{Unified approach.} We now demonstrate that the expressions for $\Delta_{\mathrm{QB}}$ and $\Delta_{\mathrm{RE}}$ with the range $r = (-\infty, 0)$ can be cast in the same mathematical form. For general backflow, we write the time-dependent wave function of a free particle as
\begin{equation}
	\Psi(x,t) = \frac{1}{\sqrt{2 \pi \hbar}} \int_{-\infty}^{+\infty} dp \, \tilde{\psi}(p) \exp \left( -\frac{i t}{2 m \hbar} p^2 + \frac{i x}{\hbar} p \right),
\end{equation}
where $\tilde{\psi}(p)$ is the normalized momentum-space wave function at $t = 0$. The relevant probabilities in Eq.~\eqref{EQ_DELTA_QB} are then
\begin{equation}
	P_-(t) = \int_{-\infty}^0 dx \, |\Psi(x,t)|^2 \,, \quad \tilde{P}_- = \int_{-\infty}^0 dp \, \big| \tilde{\psi}(p) \big|^2 \,.
\label{EQ_for_P's}
\end{equation}
Following the approach of Ref.~\cite{Bracken1994}, detailed for convenience in the SM, we find that $\Delta_{\mathrm{QB}} = \Delta$, where
\begin{eqnarray}
	\Delta & = & \int_{-\infty}^{+\infty} du \int_{-\infty}^{+\infty} du' \, \varphi^*(u) K(u, u') \varphi(u') \,,
\label{EQ_Delta} \\
	K(u,u') & = & - \frac{\sin \left( u^2 - {u'}^2 \right)}{\pi (u - u')} - \Theta(-u) \delta(u - u') \,,
\label{EQ_K}
\end{eqnarray}
and
\begin{equation}
\varphi(u) = \left( \frac{4 \hbar m}{t_2 - t_1} \right)^{\!\! 1/4} \exp \! \left( -i \frac{t_2 + t_1}{t_2 - t_1} u^2 \right) \tilde{\psi} \! \left( u \sqrt{\frac{4 \hbar m}{t_2 - t_1}} \right) \,.
\end{equation}
Here, $\delta(\cdot)$ denotes the Dirac delta function, and $\varphi(u)$ is a rescaled momentum-space wave function normalized as $\int_{-\infty}^{+\infty} du \, |\varphi(u)|^2 = 1$.

For quantum reentry with $r = (-\infty, 0)$, we express $\Psi(x,t)$ via the free-particle propagator
\begin{equation}
	\Psi(x,t) = \sqrt{\frac{m}{2 \pi i \hbar \tau}} \int_{-\infty}^{+\infty} dx' \, \exp \! \left( i \frac{m (x - x')^2}{2 \hbar \tau} \right) \psi(x') \,,
\end{equation}
where $\tau = t - t_0$, and $\psi(x)$ is the position-space wave function at $t = t_0$. Following the method of Ref.~\cite{Goussev2019}, detailed in the SM, $\Delta_{\mathrm{RE}}$ can be expressed in the same form, $\Delta_{\mathrm{RE}} = \Delta$, with $\varphi(u)$ now representing a rescaled position-space wave function:
\begin{eqnarray}
\varphi(u) & = & \left( \frac{4 \hbar \tau_2 \tau_1}{m (\tau_2-\tau_1)} \right)^{\!\! 1/4} \exp \! \left( i \frac{\tau_2+\tau_1}{\tau_2-\tau_1} u^2\right) \nonumber \\
& \times & \psi \! \left( -u \sqrt{\frac{4 \hbar \tau_2 \tau_1}{m (\tau_2-\tau_1)}} \right) \,,
\end{eqnarray}
where $\tau_1 = t_1 - t_0$ and $\tau_2 = t_2 - t_0$.

Thus, $\Delta$ captures the probability transfer in both the general backflow and reentry scenarios. Compared with the standard backflow~\cite{Bracken1994} and reentry~\cite{Goussev2019} formulations, two key differences appear: the integrals over $u$ and $u'$ extend over the entire real line, and the kernel $K(u,u')$ includes an additional term, $-\Theta(-u) \delta(u - u')$. These modifications account for the larger values of $\Delta$ observed in the general setting. In the SM we also give a perturbative argument, based on modifications of the state maximizing standard positive-momentum backflow, showing that $\Delta$ can exceed the Bracken–Melloy limit.

\emph{Supremum of $\Delta$.} The maximal quantum violation of the classical inequalities~\eqref{EQ_B} and~\eqref{EQ_RE} is obtained by optimizing $\Delta$ over all normalized functions $\varphi(u)$. This is achieved by performing an unconstrained optimization of $\Delta + \lambda \int_{-\infty}^{+\infty} du \, |\varphi(u)|^2$, where $\lambda$ is a Lagrange multiplier~\cite{Bracken1994}. The resulting Euler-Lagrange equation reads
\begin{equation}
	\int_{-\infty}^{+\infty} du' \, K(u,u') \varphi(u') = \lambda \, \varphi(u). \label{EQ_EIGEN_RESCALED}
\end{equation}
so that the supremum of $\Delta$ corresponds to the largest eigenvalue of the integral operator: $\sup_{\varphi} \Delta = \sup \{ \lambda \}$.

To evaluate $\sup_{\varphi} \Delta$ numerically, we first truncate the integration domain in Eq.~\eqref{EQ_EIGEN_RESCALED} to $(-L, L)$, giving eigenvalues $\lambda_L$ such that $\lambda = \lim_{L \to \infty} \lambda_L$. We then discretize $(-L, L)$ into $2N$ intervals, replacing the integral with a finite sum, thereby converting the integral eigenproblem into a matrix eigenproblem with eigenvalues $\lambda_{L,N}$, which satisfy $\lim_{N \to \infty} \lambda_{L,N} = \lambda_L$.

In our calculations, $L$ was varied from $10$ to $40$ in steps of $5$, and for each $L$, $N$ was increased from hundreds to thousands to determine the maximal $\lambda_{L,N}$. For fixed $L$, the maximal $\lambda_{L,N}$ scales approximately linearly with $1/N$; we therefore estimate the limit $N \to \infty$ and obtain $\lambda_L$ by fitting a linear function to the computed eigenvalues. The optimal fit parameters and associated uncertainties are obtained via standard linear regression~\cite{Bevington2003}. Similarly, the maximal $\lambda_L$ scales approximately linearly with $1/L$, allowing us to estimate the limit $L \to \infty$ and its uncertainty. The resulting value, with one standard deviation error, is
\begin{equation}
	\sup \Delta = 0.128100 \pm 0.000002 \,.
	\label{EQ_SUP_N}
\end{equation}
Comprehensive details of the numerical analysis are provided in the SM.

\begin{figure}[!t]
	\centering
	\includegraphics[width=0.49\textwidth]{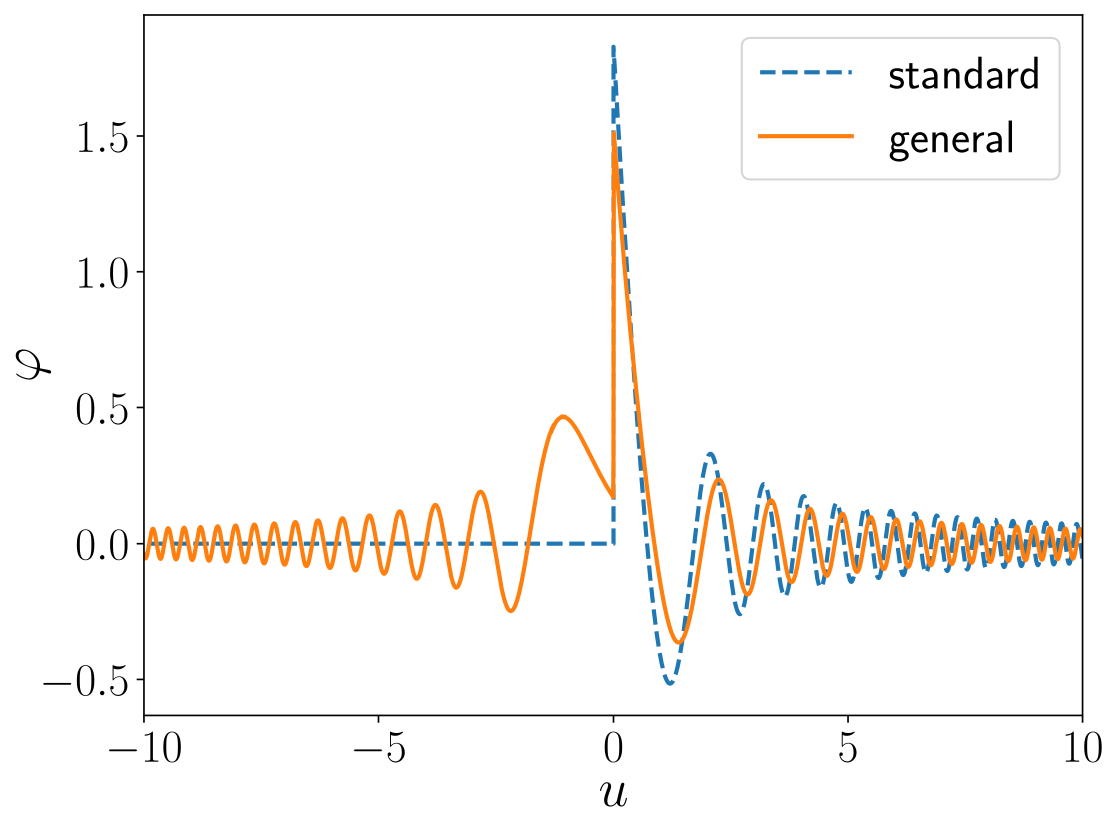}
	\caption{Rescaled state maximizing general backflow and reentry compared with the standard (positive-momentum) backflow-maximizing state.}
	\label{FIG_OPT_STATE}
\end{figure}
The eigenvector corresponding to the largest eigenvalue provides an approximation of the state $\varphi(u)$ that maximizes general backflow or reentry. As shown in Fig.~\ref{FIG_OPT_STATE} this state (solid orange curve) exhibits a discontinuity at $u = 0$ and long oscillatory tails extending to infinity, similar to the standard backflow-maximizing state (dashed blue curve for positive-momentum problem).

\emph{Analytical bounds.} Analytical lower and upper bounds on $\sup \Delta$ can be inferred from the results of Refs.~\cite{Trillo2023, Zaw2022, Zaw2025}. Let us express $\tilde{P}_- = \mathrm{Tr}(\rho \, \Theta(-\hat{p}))$, where $\rho$ is the initial density matrix of the particle and $\hat{p}$ is the momentum operator. The probability of finding the particle at negative positions at time $t$ is then $P_-(t) = \mathrm{Tr}(\rho_t \, \Theta(-\hat{x})) = \mathrm{Tr}(\rho \, \Theta(-\hat{x} - \hat{p} t / m))$, where $\hat{x}$ is the position operator, $\rho_t = U \rho U^\dagger$, and $U = \exp(-i t \hat{p}^2 / 2 \hbar m)$ is the free-evolution operator.
For the case $m = 1$, $t_1 = 0$, and $t_2 = 1$, Eq.~\eqref{EQ_DELTA_QB} can be written as $\Delta_{\mathrm{QB}} = \mathrm{Tr}(\rho \, \Omega)$, with $\Omega = \Theta(-\hat{x} - \hat{p}) - \Theta(-\hat{x}) - \Theta(-\hat{p})$. The operator $\Omega$ was introduced in Ref.~\cite{Trillo2023} to estimate the transport efficiency of quantum projectiles with the same position and momentum marginals as their classical counterparts. (These assumptions are not present in our analysis.)

The best known analytical bounds on the maximal eigenvalue of $\Omega$ were established by connecting it to Tsirelson's precession problem~\cite{Tsirelson2006}, and applying very recent results of Ref.~\cite{Zaw2022,Zaw2025}. This shows that $0.128092 \le \sup \Delta \le 0.192466$, exceeding the Bracken-Melloy bound $c_{\text{BM}} \simeq 0.0384506$ by more than a factor of three, and with the lower bound coming remarkably close to our estimate, given by Eq.~\eqref{EQ_SUP_N}.

\emph{Discussion.} We conclude with remarks on the foundational implications of our results. In the general backflow formulation, we considered the probability of locating the particle to the left of the origin. Naturally, the origin can be replaced by any point $b$ on the $x$ axis. Quantum backflow occurs whenever $\Delta_{\text{QB}} = \prob(x < b, t_2) - \prob(x < b, t_1) - \tilde{P}_-$ [c.f.~Eq.~\eqref{EQ_DELTA_QB}] becomes positive. Similarly, one can consider quantum ``overflow'', recently introduced in Ref.~\cite{FK--Repeated}, which occurs when $\Delta_{\text{QO}} = \prob(x > f, t_2) - \prob(x > f, t_1) - \tilde{P}_+$ becomes positive, with $f$ an arbitrary point and $\tilde{P}_+ = 1 - \tilde{P}_-$ the probability of positive momentum. Since in the general formulation introduced here the particle can move in any direction, a natural question arises: can a particle simultaneously exhibit backflow and overflow? In other words, does there exist a wave function for which $\Delta_{\text{QB}} > 0$ and $\Delta_{\text{QO}} > 0$ for some choice of parameters? The answer is negative. It can be shown (see SM) that $\Delta_{\mathrm{QB}} + \Delta_{\mathrm{QO}} \le 0$ for all parameters, revealing a trade-off and mutual incompatibility between these two effects. It is also worth noting that while quantum overflow can occur only over multiple disjoint time intervals in states with strictly positive momentum~\cite{FK--Repeated}, it can already arise within a single interval for states with an arbitrary momentum distributions. In fact, quantum backflow and overflow are qualitatively equivalent phenomena, distinguished only by the direction of the probability transport.

The Bracken–Melloy constant $c_{\mathrm{BM}}$ characterizes maximal standard (positive-momentum) backflow and is independent of the particle’s mass, the backflow duration, and Planck’s constant. It was proposed to reflect the mathematical structure of quantum free-particle evolution~\cite{Bracken1994}. The same properties hold for the general backflow constant $\sup \Delta$, given by Eq.~\eqref{EQ_SUP_N}, suggesting that quantum free evolution is characterized by at least two fundamental dimensionless constants. Understanding the relation between them presents a promising direction for future research.

A more pragmatic follow-up question is how the strength of general backflow changes when the experimental evaluation of $\prob(x<0,t)$ uses a smoothed spatial window rather than a sharp one. In the case of the standard positive-momentum formulation, such smoothing is known to lead to an underestimation of quantum backflow~\cite{Yearsley2012}. Another important aspect of a future experimental validation of general backflow is ensuring that any violation of inequalities \eqref{EQ_B} or \eqref{EQ_RE} arises from quantum interference rather than from a failure of the dynamical assumptions (e.g., force-free motion) of the system under investigation. A theoretical framework for eliminating such a dynamical loophole in the context of the Tsirelson precession problem has recently been developed in Ref.~\cite{Garg2025_arXiv}.

The classical limits, Eqs.~\eqref{EQ_B} and \eqref{EQ_RE}, rely on two assumptions: (i) the existence of a joint probability density $f(x,p,t)$, and (ii) the validity of classical equations of motion for a free particle, $x_t = x_0 + pt/m$ and $p_t = \text{const}$. Quantum backflow, reentry, and other nonclassical probability transport effects violate at least one of these assumptions. Relaxing (i) while keeping (ii) leads to pseudo-probability frameworks, such as the Wigner function formalism, in which backflow emerges as negative probability transfer of particles in straight line motion~\cite{Bracken1994,Yearsley2012}. Conversely, retaining (i) and relaxing (ii) leads to Bohmian mechanics, where even free particles follow curved trajectories due to an effective nonlocal potential~\cite{ML00Arrival, PGKW06new}. Yet, these trajectories have to comply with the Ehrenfest theorem. Moreover, one can demonstrate that any covariance measurements remain compatible with the straight line motion. Consequently, any signatures of backflow can only appear in the dynamics of third and higher moments of the position distribution. This demonstrates both the subtlety of the effect and the sensitivity of our methods to these higher-order changes.

We also note that the derivation of the inequalities \eqref{EQ_B} and \eqref{EQ_RE} does not require strictly free-particle equations of motion. Rather, they hold for any continuous phase-space flow satisfying $\sgn(d x_t / d t) = \sgn(p_t)$ and $\sgn(p_t) = \text{const}$, where $\sgn(\cdot)$ denotes the sign function. These conditions require that the particle's velocity be aligned with its momentum and that the direction of motion be preserved. This observation opens the possibility of investigating general quantum backflow and reentry beyond the free-particle setting.

In summary, the methods developed here provide a realistic route toward experimental observation of quantum backflow and reentry, while raising several fundamental and practical questions for further exploration.

\emph{Acknowledgments.} The authors thank Anat Daniel, Bohnishikha Ghosh, Bernard Gorzkowski, Thomas Juffmann, Radosław Łapkiewicz, Timon Zipfelmaier for insightful discussions.

%%%%%%%%%%%%%%%%%%%%%%%%%%%%%%%%%%%%
%\bibliography{cite}
%%%%%%%%%%%%%%%%%%%%%%%%%%%%%%%%%%%%

%

\begin{widetext}

\section{Supplemental Material}
  
\subsection{Alternative derivation of classical constraints}

\begin{figure}[!b]
\includegraphics[width=0.8\textwidth]{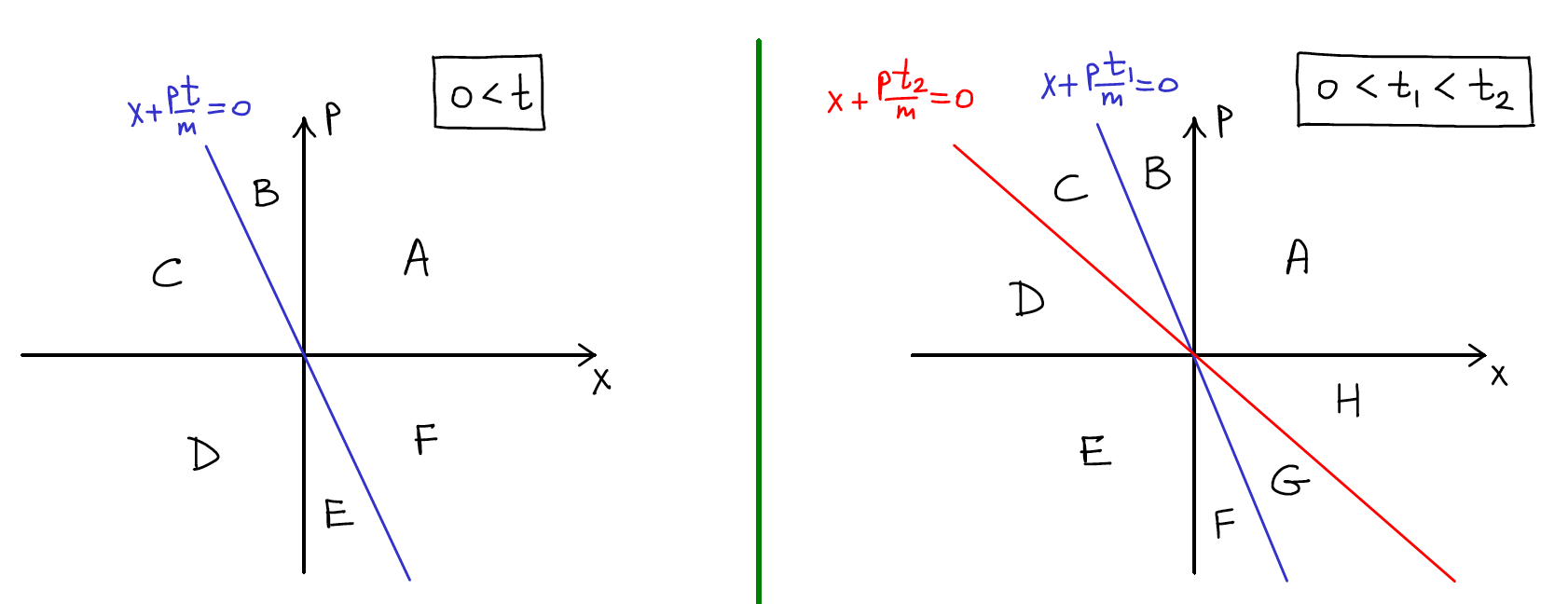}
\caption{\label{FIG_INEQ} Phase-space diagram used to derive inequalities characterizing the absence of backflow and reentry in classical mechanics.}
\end{figure}

Consider the left panel of Fig.~\ref{FIG_INEQ}, which illustrates the phase space of classical one-dimensional motion. A classical particle is measured first at time $t = 0$ and then again at time $t$. For free particles, the phase-space flow is purely horizontal, mapping lines through the origin to other lines through the origin. This leads to the following regions that contribute to the marginal probabilities:
\begin{eqnarray*}
    P_{-}(0) & = & \int_{B+C+D} dp \, dx \, f(x,p,0) \,, \\
    P_{-}(t) & = & \int_{C+D+E} dp \, dx \, f(x,p,0) \,, \\
    \tilde P_{-} & = & \int_{D+E+F} dp \, dx \, f(x,p,0) \,.
\end{eqnarray*}
These can be treated as follows:
\begin{eqnarray*}
    P_{-}(t) & \le & \int_{C+D+E} dp \, dx \, f(x,p,0) + \int_{B} dp \, dx \, f(x,p,0) = P_{-}(0) + \int_{E} dp \, dx \, f(x,p,0) \\
     & \le & P_{-}(0) + \int_{D+E+F} dp \, dx \, f(x,p,0) = P_{-}(0) + \tilde P_{-} \,.
\end{eqnarray*}
This inequality characterizes the absence of backflow.

The right panel of Fig.~\ref{FIG_INEQ} illustrates the scenario in which only position measurements are performed at three distinct instants of time. The corresponding probabilities are given by
\begin{eqnarray*}
    P_{-}(t_1) & = & \int_{C+D+E+F} dp \, dx \, f(x,p,0) \,, \\
    P_{-}(t_2) & = & \int_{D+E+F+G} dp \, dx \, f(x,p,0) \,, \\
    1-P_{-}(0) & = & \int_{A+H+G+F} dp \, dx \, f(x,p,0) \,.
\end{eqnarray*}
Therefore,
\begin{eqnarray*}
    P_{-}(t_2) - P_{-}(t_1) & = & \int_{G} dp \, dx \, f(x,p,0) - \int_{C} dp \, dx \, f(x,p,0) \\
     & \le & \int_{G+A+F+H} dp \, dx \, f(x,p,0) = 1 - P_{-}(0) \,.
\end{eqnarray*}
This inequality characterizes the absence of reentry in classical systems.

%%%%%%%%%%%%%%%%%%%%%%%%%%%%%%%%%%%%
\subsection{Rescaling of general backflow}
\label{sec:backflow}
%%%%%%%%%%%%%%%%%%%%%%%%%%%%%%%%%%%%

The time-dependent wave function of a free particle can be written as
\begin{equation*}
	\Psi(x,t) = \frac{1}{\sqrt{2 \pi \hbar}} \int_{-\infty}^{+\infty} dp \, \tilde{\psi}(p) \exp \! \left( -\frac{i t}{2 m \hbar} p^2 + \frac{i x}{\hbar} p \right) \,,
\end{equation*}
where $\tilde{\psi}(p)$ is the momentum-space wave function at time $t=0$. The latter satisfies the normalization condition
\begin{equation*}
	\int_{-\infty}^{+\infty} dp \, \big| \tilde{\psi}(p) \big|^2 = 1 \,.
\end{equation*}
The associated probability current through $x=0$ at time $t$ is given by
\begin{align*}
	j(0,t)
	&= \frac{\hbar}{m} \left. \imag \! \left\{ \Psi^*(x,t) \frac{\partial \Psi(x,t)}{\partial x} \right\} \right|_{x = 0} \\[0.2cm]
	&= \frac{1}{4 \pi \hbar m} \int_{-\infty}^{+\infty} dp \int_{-\infty}^{+\infty} dp' \, \left( p + p' \right) \tilde{\psi}(p)^* \tilde{\psi} \left( p' \right) \exp \! \left[ \frac{i t}{2 \hbar m} \left( p^2 - {p'}^2 \right) \right] \,.
\end{align*}
The probability transfer into the region $x < 0$ during a time interval $(t_1, t_2)$, with $t_1 < t_2$, is
\begin{align*}
	P_-(t_2) - P_-(t_1) &= -\int_{t_1}^{t_2} dt \, j(0,t) \\
	&= -\frac{1}{4 \pi \hbar m} \int_{-\infty}^{+\infty} dp \int_{-\infty}^{+\infty} dp' \, \left( p + p' \right) \tilde{\psi}(p)^* \tilde{\psi} \left( p' \right) \int_{t_1}^{t_2} dt \, \exp \! \left[ \frac{i t}{2 \hbar m} \left( p^2 - {p'}^2 \right) \right] \\
	&= -\frac{1}{2 \pi i} \int_{-\infty}^{+\infty} dp \int_{-\infty}^{+\infty} dp' \, \left( p + p' \right) \tilde{\psi}(p)^* \tilde{\psi} \left( p' \right) \frac{\exp \! \left[ \frac{i t_2}{2 \hbar m} ( p^2 - {p'}^2 ) \right] - \exp \! \left[ \frac{i t_1}{2 \hbar m} ( p^2 - {p'}^2 ) \right]}{p^2 - p'^2} \,.
\end{align*}
Writing
\begin{equation*}
	t_1 = \bar{t} - \frac{T}{2} \,, \quad t_2 = \bar{t} + \frac{T}{2} \,,
\end{equation*}
we obtain
\begin{equation*}
	P_-(t_2) - P_-(t_1) = -\frac{1}{\pi} \int_{-\infty}^{+\infty} dp \int_{-\infty}^{+\infty} dp' \, \tilde{\psi}(p)^* \tilde{\psi}(p') \exp \! \left[ \frac{i \bar{t}}{2 \hbar m} ( p^2 - {p'}^2 ) \right] \frac{\sin \! \left[ \frac{T}{4 \hbar m} ( p^2 - {p'}^2 ) \right]}{p - p'} \,.
\end{equation*}
Changing the integration variables to
\begin{equation*}
	u = p \sqrt{\frac{T}{4 \hbar m}} \,, \quad u' = p' \sqrt{\frac{T}{4 \hbar m}} \,,
\end{equation*}
we get
\begin{equation*}
	P_-(t_2) - P_-(t_1) = -\frac{1}{\pi} \int_{-\infty}^{+\infty} du \int_{-\infty}^{+\infty} du' \, \sqrt{\frac{4 \hbar m}{T}} \, \tilde{\psi} \! \left( u \sqrt{\frac{4 \hbar m}{T}} \right)^{\!\! *} \tilde{\psi} \! \left( u' \sqrt{\frac{4 \hbar m}{T}} \right) \exp \! \left[ i \frac{2 \bar{t}}{T} \! \left( u^2 - u'^2 \right) \right] \frac{\sin(u^2 - u'^2)}{u - u'} \,.
\end{equation*}
We now introduce the rescaled momentum-space wave function
\begin{align*}
	\varphi(u) &= \left( \frac{4 \hbar m}{T} \right)^{\!\! 1/4} \exp \! \left( -i \frac{2 \bar{t}}{T} u^2 \right) \tilde{\psi} \! \left( u \sqrt{\frac{4 \hbar m}{T}} \right) \\
	&= \left( \frac{4 \hbar m}{t_2 - t_1} \right)^{\!\! 1/4} \exp \! \left( -i \frac{t_2 + t_1}{t_2 - t_1} u^2 \right) \tilde{\psi} \! \left( u \sqrt{\frac{4 \hbar m}{t_2 - t_1}} \right) \,.
\end{align*}
Then,
\begin{equation*}
	P_-(t_2) - P_-(t_1) = -\frac{1}{\pi} \int_{-\infty}^{+\infty} du \int_{-\infty}^{+\infty} du' \, \varphi(u)^* \varphi(u') \frac{\sin(u^2 - u'^2)}{u - u'} \,.
\end{equation*}
The normalization condition now reads
\begin{equation*}
	\int_{-\infty}^{+\infty} du \, | \varphi(u) |^2 = 1 \,,
\end{equation*}
and the probability of negative momentum becomes
\begin{equation*}
	\tilde{P}_- = \int_{-\infty}^0 du \, | \varphi(u) |^2 \,.
\end{equation*}
Combining the results, we find
\begin{align*}
	\Delta_{\text{QB}} &= P_-(t_2) - P_-(t_1) - \tilde{P}_- \\
	&= -\frac{1}{\pi} \int_{-\infty}^{+\infty} du \int_{-\infty}^{+\infty} du' \, \varphi(u)^* \varphi(u') \frac{\sin(u^2 - u'^2)}{u - u'} - \int_{-\infty}^0 du \, | \varphi(u) |^2 \\
	&= -\frac{1}{\pi} \int_{-\infty}^{+\infty} du \int_{-\infty}^{+\infty} du' \, \varphi(u)^* \varphi(u') \frac{\sin(u^2 - u'^2)}{u - u'} - \int_{-\infty}^{+\infty} du \int_{-\infty}^{+\infty} du' \, \varphi(u)^* \varphi(u') \Theta(-u) \delta(u - u') \\
	&= \int_{-\infty}^{+\infty} du \int_{-\infty}^{+\infty} du' \, \varphi(u)^* K(u,u') \varphi(u') \,,
\end{align*}
where
\begin{equation*}
	K(u,u') = -\frac{1}{\pi} \frac{\sin(u^2 - u'^2)}{u - u'} - \Theta(-u) \delta(u - u') \,.
\end{equation*}

%%%%%%%%%%%%%%%%%%%%%%%%%%%%%%%%%%%%
\subsection{Rescaling of general reentry}
\label{sec:reentry}
%%%%%%%%%%%%%%%%%%%%%%%%%%%%%%%%%%%%

The time-dependent wave function of a free particle can also be expressed via the free-particle propagator:
\begin{equation*}
	\Psi(x,t) = \sqrt{\frac{m}{2 \pi i \hbar \tau}} \int_{-\infty}^{+\infty} dx' \, \exp \! \left( i \frac{m (x - x')^2}{2 \hbar \tau} \right) \psi(x') \,,
\end{equation*}
where $\tau = t - t_0$, and $\psi(x)$ is the position-space wave function at $t = t_0$ that satisfies the normalization condition
\begin{equation*}
	\int_{-\infty}^{+\infty} dx \, |\psi(x)|^2 = 1 \,.
\end{equation*}
In this representation, the probability current through $x=0$ at time $t$ becomes
\begin{align*}
	j(0,t)
	&= \frac{\hbar}{m} \left. \imag \! \left\{ \Psi^*(x,t) \frac{\partial \Psi(x,t)}{\partial x} \right\} \right|_{x = 0} \\[0.2cm]
	&= -\frac{m}{4 \pi \hbar \tau^2} \int_{-\infty}^{+\infty} dx \int_{-\infty}^{+\infty} dx' \, (x + x') \exp \! \left[ -i \frac{m \left( x^2 - x'^2 \right)}{2 \hbar \tau} \right] \psi(x)^* \psi(x') \,.
\end{align*}
The probability transfer into the region $x < 0$ during a time interval $(t_1, t_2)$, with $t_1 < t_2$, is
\begin{align*}
	P_-(t_2) - P_-(t_1) &= -\int_{t_1}^{t_2} dt \, j(0,t) \\
	&= \frac{m}{4 \pi \hbar} \int_{-\infty}^{+\infty} dx \int_{-\infty}^{+\infty} dx' \, (x + x') \psi(x)^* \psi(x') \int_{t_1 - t_0}^{t_2 - t_0} \frac{d\tau}{\tau^2} \exp \! \left[ -i \frac{m \left( x^2 - x'^2 \right)}{2 \hbar \tau} \right] \,.
\end{align*}
Making the substitution $z = 1/\tau$, we write
\begin{align*}
	P_-(t_2) - P_-(t_1) &= \frac{m}{4 \pi \hbar} \int_{-\infty}^{+\infty} dx \int_{-\infty}^{+\infty} dx' \, (x + x') \psi(x)^* \psi(x') \int_{1/(t_2 - t_0)}^{1/(t_1 - t_0)} dz \, \exp \! \left[ -i \frac{m \left( x^2 - x'^2 \right)}{2 \hbar} z \right] \\
	&= \frac{i}{2 \pi} \int_{-\infty}^{+\infty} dx \int_{-\infty}^{+\infty} dx' \, (x + x') \psi(x)^* \psi(x') \frac{\exp \! \left[ -i \frac{m \left( x^2 - x'^2 \right)}{2 \hbar (t_1-t_0)} \right] - \exp \! \left[ -i \frac{m \left( x^2 - x'^2 \right)}{2 \hbar (t_2 - t_0)} \right]}{x^2 - x'^2} \,.
\end{align*}
Introducing $\nu$ and $\omega$ such that
\begin{equation*}
	\frac{1}{t_2-t_0} = \nu - \frac{\omega}{2} \,, \quad \frac{1}{t_1-t_0} = \nu + \frac{\omega}{2} \,,
\end{equation*}
and, consequently,
\begin{equation*}
	\nu = \frac{t_2 + t_1 - 2 t_0}{2 (t_2-t_0) (t_1-t_0)} \,, \quad \omega = \frac{t_2 - t_1}{(t_2-t_0) (t_1-t_0)} \,,
\end{equation*}
we get
\begin{align*}
	P_-(t_2) - P_-(t_1)
	&= \frac{i}{2 \pi} \int_{-\infty}^{+\infty} dx \int_{-\infty}^{+\infty} dx' \, \psi(x)^* \psi(x') \exp \! \left[ -i \frac{m \left( x^2 - x'^2 \right)}{2 \hbar} \nu \right] \frac{\exp \! \left[ -i \frac{m \left( x^2 - x'^2 \right)}{4 \hbar} \omega \right] - \exp \! \left[ i \frac{m \left( x^2 - x'^2 \right)}{4 \hbar} \omega \right]}{x - x'} \\
	&= \frac{1}{\pi} \int_{-\infty}^{+\infty} dx \int_{-\infty}^{+\infty} dx' \, \psi(x)^* \psi(x') \exp \! \left[ -i \frac{m \left( x^2 - x'^2 \right)}{2 \hbar} \nu \right] \frac{\sin \! \left[ \frac{m \left( x^2 - x'^2 \right)}{4 \hbar} \omega \right]}{x-x'} \,.
\end{align*}
We then change the integration variables as
\begin{align*}
	u = -x \sqrt{\frac{m \omega}{4 \hbar}} \,, \quad u' = -x' \sqrt{\frac{m \omega}{4 \hbar}} \,.
\end{align*}
This yields
\begin{equation*}
	P_-(t_2) - P_-(t_1)	= -\frac{1}{\pi} \int_{-\infty}^{+\infty} du \int_{-\infty}^{+\infty} du' \, \sqrt{\frac{4 \hbar}{m \omega}} \psi \! \left( -u \sqrt{\frac{4 \hbar}{m \omega}} \right)^{\!\! *} \psi \! \left( -u' \sqrt{\frac{4 \hbar}{m \omega}} \right) \exp \! \left[ -i \frac{2 \nu}{\omega} \left( u^2 - u'^2 \right) \right] \frac{\sin(u^2 - u'^2)}{u-u'} \,.
\end{equation*}
Now, defining
\begin{align*}
	\varphi(u) &= \left( \frac{4 \hbar}{m \omega} \right)^{\!\! 1/4} \exp \! \left( i \frac{2 \nu}{\omega} u^2\right) \psi \! \left( -u \sqrt{\frac{4 \hbar}{m \omega}} \right) \\
	&= \left( \frac{4 \hbar (t_2-t_0) (t_1-t_0)}{m (t_2-t_1)} \right)^{\!\! 1/4} \exp \! \left( i \frac{t_2+t_1-2t_0}{t_2-t_1} u^2\right) \psi \! \left( -u \sqrt{\frac{4 \hbar (t_2-t_0) (t_1-t_0)}{m (t_2-t_1)}} \right) \,,
\end{align*}
we obtain
\begin{equation*}
	P_-(t_2) - P_-(t_1)	= -\frac{1}{\pi} \int_{-\infty}^{+\infty} du \int_{-\infty}^{+\infty} du' \, \varphi(u)^* \varphi(u') \frac{\sin(u^2 - u'^2)}{u-u'} \,.
\end{equation*}
In view of the normalization of $\psi(x)$, the rescaled wave function $\varphi(u)$ is normalized to unity,
\begin{equation*}
	\int_{-\infty}^{+\infty} du \, | \varphi(u) |^2 = 1 \,.
\end{equation*}
Finally, since
\begin{align*}
	1 - P_-(t_0) &= \int_0^{+\infty} dx \, |\psi(x)|^2 \\
	&= \int_{-\infty}^0 du \, |\varphi(u)|^2 \\
	&= \int_{-\infty}^{+\infty} du \int_{-\infty}^{+\infty} du' \, \Theta(-u) \delta(u-u') \varphi(u)^* \varphi(u') \,,
\end{align*}
we obtain
\begin{align*}
	\Delta_{\text{RE}} &= P_-(t_2) - P_-(t_1) - \big( 1 - P_-(t_0) \big) \\
	&= \int_{-\infty}^{+\infty} du \int_{-\infty}^{+\infty} du' \, \varphi(u)^* K(u,u') \varphi(u') \,,
\end{align*}
where
\begin{equation*}
	K(u,u') = -\frac{1}{\pi} \frac{\sin(u^2 - u'^2)}{u - u'} - \Theta(-u) \delta(u - u') \,.
\end{equation*}

%%%%%%%%%%%%%%%%%%%%%%%%%%%%%%%%%%%%
\subsection{Superposition of two Gaussians}
%%%%%%%%%%%%%%%%%%%%%%%%%%%%%%%%%%%%

\begin{figure}
    \centering
    \includegraphics[width=0.6\linewidth]{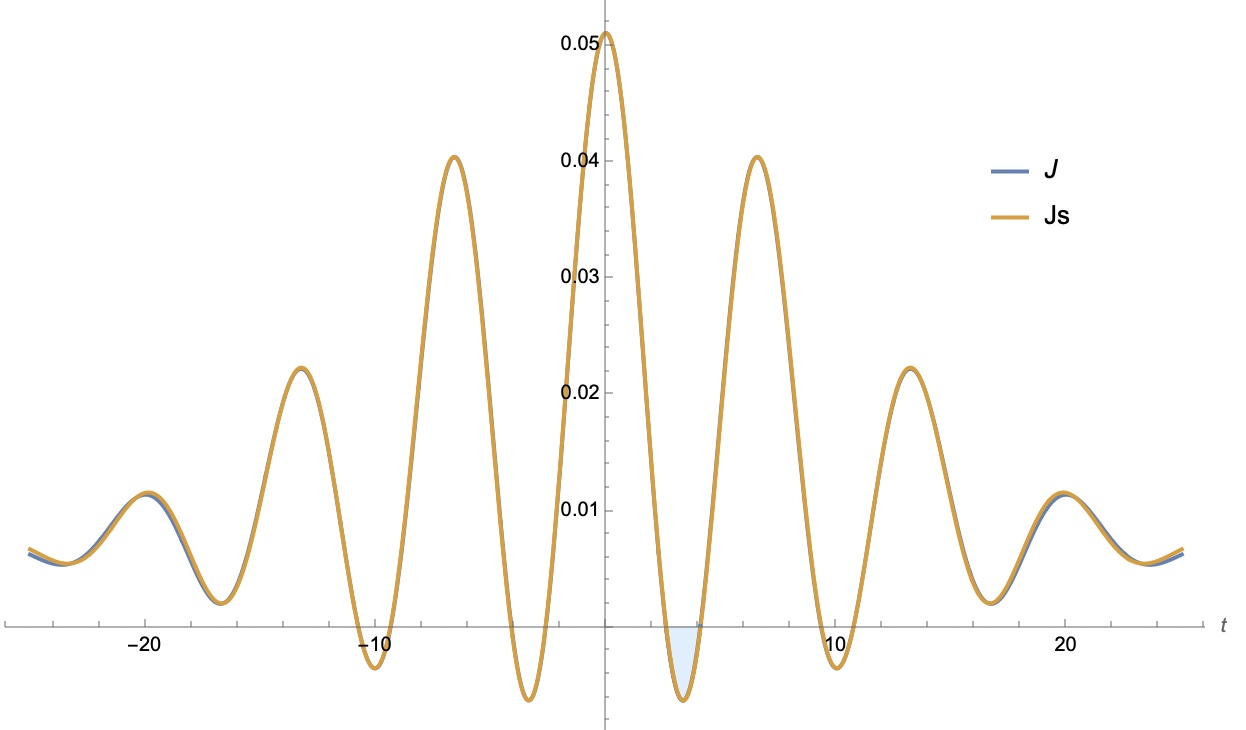}
    \caption{Approximation to probability current in standard backflow. The plot shows the probability current, label $J$, and the simple approximation given by Eq.~(\ref{EQ_SIM_J}), label $J_s$, for the optimal parameters of Ref.~\cite{Yearsley2012}. The backflow interval is marked with light blue area. The two curves are practically indistinguishable up to $t \simeq 15$ units.}
    \label{FIG_YEARSLEY}
\end{figure}

For comparison with the standard backflow, we first derive a simple approximation to the probability current. 
Consider the following quantum state at time $t$:
\begin{eqnarray*}
\psi(x,t) = c_1 g_1(x,t) + c_2 g_2(x,t) \,,
\end{eqnarray*}
where $|c_1|^2 + |c_2|^2 = 1$, and the free-particle Gaussian wave functions read
\begin{equation*}
g_n(x,t) = \left( \frac{1}{2 \pi \sigma_n^2 (1 + i \omega_n t)^2} \right)^{1/4} \exp\left(- \frac{(x - p_n t / m)^2}{4 \sigma_n^2 (1 + i \omega_n t)} \right) \exp\left(\frac{i}{\hbar} p_n x \right) \exp\left( - \frac{i t}{\hbar} \frac{p_n^2}{2m} \right) \,, \qquad
\omega_n = \frac{\hbar}{2 m \sigma_n^2} \,.
\end{equation*}
Here, $p_n$ is the mean momentum of the $n$th Gaussian, and $\sigma_n$ is its standard deviation.
The probability current at $x=0$ and time $t$ can be written as
\begin{equation*}
j(0,t) = j_1(0,t) + j_2(0,t) + j_{\cross}(0,t) \,,
\end{equation*}
with the contributions from the individual Gaussians given by
\begin{equation*}
j_n(0,t) = \frac{p_n}{m} |c_n|^2 |g_n(0,t)|^2 \frac{1}{1+\omega_n^2 t^2} \,,
\end{equation*}
and the cross-term:
\begin{eqnarray*}
j_{\cross} (0,t) & = & \frac{|z|}{m} \cos\varphi \left( \frac{p_2}{1 + \omega_2^2 t^2} + \frac{p_1}{1 + \omega_1^2 t^2}\right)
+\frac{|z|}{m} \sin \varphi \left( \frac{p_2 \omega_2 t}{1 + \omega_2^2 t^2} - \frac{p_1 \omega_1 t}{1 + \omega_1^2 t^2}\right),
\end{eqnarray*}
where $|z| e^{i \varphi} = c_1^* c_2 g_1(0,t)^* g_2(0,t)$.

For comparison with Ref.~\cite{Yearsley2012} we assume that $\sigma_1 = \sigma_2 \equiv \sigma$, hence $\omega_1 = \omega_2 \equiv \omega$, and that coefficients $c_1$ and $c_2$ are purely real.
In this case,
\begin{eqnarray*}
|z| & = & c_1 c_2 \left( \frac{1}{2 \pi \sigma^2(1 + \omega^2 t^2)} \right)^{1/2} \exp \left(- \frac{(p_1 t/m)^2 + (p_2 t/m)^2}{4 \sigma^2(1+\omega^2 t^2)} \right) \,, \\
\varphi & = & \frac{(p_1^2 - p_2^2) t}{2 \hbar m (1 + \omega^2 t^2)} \,.
\end{eqnarray*}
The current admits a simple form under further assumption that $\omega t \ll 1$. It turns out that this is not restrictive and includes maximal standard backflow identified in \cite{Yearsley2012}, as shown in Fig.~\ref{FIG_YEARSLEY}. The current reads:
\begin{eqnarray}
    j_s = c_1^2 \, \frac{p_1}{m} \left( \frac{1}{2 \pi \sigma^2} \right)^{1/2} \exp \left(- \frac{(p_1 t/m)^2}{2 \sigma^2} \right)
    +
    c_2^2 \, \frac{p_2}{m} \left( \frac{1}{2 \pi \sigma^2} \right)^{1/2} \exp \left(- \frac{(p_2 t/m)^2}{2 \sigma^2} \right) \nonumber \\
    + c_1 c_2 \, \frac{p_1 + p_2}{m} \left( \frac{1}{2 \pi \sigma^2} \right)^{1/2} \exp \left(- \frac{(p_2 t/m)^2 + (p_1 t/m)^2}{4 \sigma^2} \right) \cos\left( \frac{(p_1^2 - p_2^2) t}{2m}  \right) \,.
    \label{EQ_SIM_J}
\end{eqnarray}
Using these assumptions on the parameters, we numerically explored the general backflow quantity $\Delta_{\mathrm{QB}} = -\int_{t_1}^{t_2} dt \, j(0,t) - \tilde{P}_-$ and found the quoted value of $0.0106$ attained for $c_1 = 1.8$, $c_2 = 1$, $\sigma = 1.6$, $p_1 = 1$, $p_2 = 5.7$, $t_1 = 0.152246$, and $t_2 = 0.246813$.

For the full optimization of $\Delta_{\mathrm{QB}}$, as well as of $\Delta_{\mathrm{RE}}$, over the most general superposition of two Gaussian packets (without the assumption considered above), we write the problem in the dimensionless form:
\begin{equation*}
	\Delta = -\frac{1}{\pi} \int_{-\infty}^{+\infty} du \int_{-\infty}^{+\infty} du' \, \varphi(u)^* \frac{\sin(u^2 - u'^2)}{u - u'} \varphi(u') - \int_{-\infty}^0 du \, |\varphi(u)|^2
\end{equation*}
with
\begin{equation*}
	\varphi(u) = C \Big( e^{-a_1 (u - b_1)^2} \cos \alpha + e^{-a_2 (u - b_2)^2} e^{i \beta} \sin \alpha \Big) \,.
\end{equation*}
Here, $a_1, b_1, a_2, b_2 \in \mathbb{C}$ and $\alpha, \beta \in \mathbb{R}$ are parameters, and the constant $C$, given by
\begin{align*}
	C^{-2} &= \cos^2 \alpha \sqrt{\frac{\pi}{2 \real a_1}} \exp \left[ \frac{2 (\real a_1 b_1)^2}{\real a_1} - 2 (\real a_1 b_1^2) \right] \\
	&\quad + \sin^2 \alpha \sqrt{\frac{\pi}{2 \real a_2}} \exp \left[ \frac{2 (\real a_2 b_2)^2}{\real a_2} - 2 (\real a_2 b_2^2) \right] \\
	&\quad + \sin 2 \alpha \real \left\{ \sqrt{\frac{\pi}{a_1 + a_2^*}} \exp \left[ \frac{(a_1 b_1 + a_2^* b_2^*)^2}{a_1 + a_2^*} - a_1 b_1^2 - (a_2 b_2^2)^* - i \beta\right] \right\} \,,
\end{align*}
ensures the normalization condition $\int_{-\infty}^{+\infty} du \, |\varphi(u)|^2 = 1$. The function $\varphi(u)$ represents the rescaled momentum-space wave function of the particle in the case of general backflow (see Sec.~\ref{sec:backflow}), and the rescaled position-space wave function in the case of general reentry (see Sec.~\ref{sec:reentry}). Performing a numerical optimization over the space of parameters $a_1, b_1, a_2, b_2, \alpha, \beta$, we find a (local) maximum with $\Delta = 0.012011$, attained for $a_1 = 28.95$, $b_1 = 0.8479$, $a_2 = 83.28$, $b_2 = 0.1637$, $\alpha = 1.271$, and $\beta = \pi$.

%%%%%%%%%%%%%%%%%%%%%%%%%%%%%%%%%%%%%%%%%%%%%%%%%%%%%%%%
\subsection{Perturbative argument}
%%%%%%%%%%%%%%%%%%%%%%%%%%%%%%%%%%%%%%%%%%%%%%%%%%%%%%%%

Here we construct a state $\varphi(u)$ that satisfies the normalization condition
\begin{equation*}
	\int_{-\infty}^{+\infty} du \, |\varphi(u)|^2 = 1
\end{equation*}
and is accompanied by
\begin{equation*}
	\Delta = -\frac{1}{\pi} \int_{-\infty}^{+\infty} du \int_{-\infty}^{+\infty} dv \, \varphi^*(u) \frac{\sin(u^2 - v^2)}{u - v} \varphi(v) - \int_{-\infty}^0 du \, |\varphi(u)|^2
\end{equation*}
larger than the Bracken-Melloy constant $c_{\text{BM}} \simeq 0.03845$.

Consider two real-valued functions $f,g : [0,+\infty) \to \mathbb{R}$, and let $\varphi(u)$ be given by
\begin{equation*}
	\varphi(u) = \left\{
	\begin{array}{ll}
		f(u) & \text{if } u > 0 \\[0.1cm]
		g(-u) & \text{if } u < 0
	\end{array} \right.
\end{equation*}
The normalization condition takes the form
\begin{equation}
	\int_0^{\infty} du \, f(u)^2 + \int_0^{\infty} du \, g(u)^2 = 1 \,,
\label{norm_fg}
\end{equation}
and $\Delta$ reads
\begin{align}
	\Delta &= -\frac{1}{\pi} \int_0^{\infty} du \int_0^{\infty} dv \, f(u) \frac{\sin(u^2 - v^2)}{u - v} f(v) \nonumber \\[0.2cm]
	&\quad + \frac{1}{\pi} \int_0^{\infty} du \int_0^{\infty} dv \, g(u) \frac{\sin(u^2 - v^2)}{u - v} g(v) \nonumber \\[0.2cm]
	&\quad - \frac{2}{\pi} \int_0^{\infty} du \int_0^{\infty} dv \, f(u) \frac{\sin(u^2 - v^2)}{u + v} g(v) - \int_0^{\infty} du \, g(u)^2 \,.
\label{Delta_fg}
\end{align}
Consider $F(u)$ to be the backflow-maximizing state in the standard Bracken-Melloy problem:
\begin{equation*}
	-\frac{1}{\pi} \int_0^{\infty} du \int_0^{\infty} dv \, F(u) \frac{\sin(u^2 - v^2)}{u - v} F(v) = c_{\text{BM}} \,, \qquad \int_0^{\infty} du \, F(u)^2 = 1 \,.
\end{equation*}
We superpose it with yet-to-be-specified function $G(u)$ such that
\begin{equation*}
	\int_0^{\infty} du \, G(u)^2 = 1 \,,
\end{equation*}
as follows:
\begin{equation*}
	f(u) = \sqrt{1 - \epsilon^2} F(u) \,, \qquad g(u) = \epsilon G(u) \,,
\end{equation*}
where $0 \le \epsilon \ll 1$. Notice that the normalization condition, Eq.~\eqref{norm_fg}, is automatically fulfilled. Equation~\eqref{Delta_fg} now reads
\begin{align*}
	\Delta &= (1 - \epsilon^2) c_{\text{BM}} \\[0.2cm]
	&\quad + \frac{\epsilon^2}{\pi} \int_0^{\infty} du \int_0^{\infty} dv \, G(u) \frac{\sin(u^2 - v^2)}{u - v} G(v) \\[0.2cm]
	&\quad - \frac{2 \epsilon \sqrt{1 - \epsilon^2}}{\pi} \int_0^{\infty} du \int_0^{\infty} dv \, F(u) \frac{\sin(u^2 - v^2)}{u + v} G(v) - \epsilon^2 \,,
\end{align*}
or
\begin{equation*}
	\Delta = c_{\text{BM}} + \epsilon I + O(\epsilon^2) \,,
\end{equation*}
where
\begin{align*}
	I
	&= \int_0^{\infty} dv H(v) G(v) \qquad \text{with} \qquad H(v) \equiv -\frac{2}{\pi} \int_0^{\infty} du \, F(u) \frac{\sin(u^2 - v^2)}{u + v} \,.
\end{align*}
The function $G(u)$ can be always chosen such that $I > 0$. Subsequently, the positivity of $I$ implies that $\Delta > c_{\text{BM}}$ for a sufficiently small $\epsilon$.

%%%%%%%%%%%%%%%%%%%%%%%%%%%%%%%%%%%%%%%%%%%%%%%%%%%%%%%%
\subsection{Example of large general backflow}
%%%%%%%%%%%%%%%%%%%%%%%%%%%%%%%%%%%%%%%%%%%%%%%%%%%%%%%%

Here we present a state that admits general backflow (or reentry) exceeding the Bracken-Melloy constant. We employ the unified framework described in Secs.~\ref{sec:backflow} and \ref{sec:reentry}, in which both $\Delta_{\text{QB}}$ and $\Delta_{\text{RE}}$ are represented by
\begin{equation*}
	\Delta = -\frac{1}{\pi} \int_{-\infty}^{+\infty} du \int_{-\infty}^{+\infty} du' \, \varphi(u)^* \frac{\sin(u^2 - u'^2)}{u - u'} \varphi(u') - \int_{-\infty}^0 du \, |\varphi(u)|^2 \,,
\end{equation*}
where $\varphi(u)$ denotes the rescaled momentum-space wave function of the particle in the case of general backflow, and the rescaled position-space wave function in the case of general reentry.

Motivated by Ref.~\cite{HGL+13Quantum}, we consider the state
\begin{equation*}
	\varphi(u) = \left\{
	\begin{array}{ll}
		C \cos \alpha \, (\beta_1 - u) e^{-\gamma_1^2 u^2} \quad &\text{if} \quad u > 0 \\[0.2cm]
		C \sin \alpha \, (\beta_2 - u) e^{-\gamma_2^2 u^2} \quad &\text{if} \quad u < 0
	\end{array} \right.
\end{equation*}
where $\alpha, \beta_1, \gamma_1, \beta_2, \gamma_2$ are real parameters, and $C$ is the normalization constant ensuring $\int_{-\infty}^{+\infty} du \, |\varphi(u)|^2 = 1$. By computing both $C$ and $\Delta$ numerically, we perform a numerical optimization of $\Delta$ over the parameter space and find a (local) maximum of $\Delta = 0.0624188$, attained for $\alpha = 0.1654$, $\beta_1 = 0.5641$, $\gamma_1 = 1.265$, $\beta_2 = 0.03529$, and $\gamma_2 = 0.6365$.

%%%%%%%%%%%%%%%%%%%%%%%%%%%%%%%%%%%%%%%%%%%%%%%%%%%%%%%%
\subsection{Backflow and overflow}
%%%%%%%%%%%%%%%%%%%%%%%%%%%%%%%%%%%%%%%%%%%%%%%%%%%%%%%%

Recall the parameter characterizing general backflow for any point $b$ on the $x$ axis:
\begin{equation}
\label{APP_D_QB}
\Delta_{\text{QB}} = \mathrm{Prob}(x < b, t_2) - \mathrm{Prob}(x < b, t_1) - \tilde{P}_-\,.\end{equation}
Similarly, we consider the parameter quantifying overflow by
\begin{equation}
\label{APP_D_QF}
\Delta_{\text{QO}} = \mathrm{Prob}(x > f, t_2) - \mathrm{Prob}(x > f, t_1) - \tilde{P}_+ \,,
\end{equation}
where $f$ is an arbitrary point and $\tilde{P}_+ = 1 - \tilde{P}_-$ is the probability of positive momentum.
We now show that both of these quantities cannot be simultaneously positive and, in fact, the following trade-off relation holds:
\begin{equation*}
\Delta_{\mathrm{QB}} + \Delta_{\mathrm{QO}} \le 0 \,.
\end{equation*}
Indeed, summing up Eqs.~(\ref{APP_D_QB}) and (\ref{APP_D_QF}), we note that the momentum probabilities add to $-1$, and the remaining terms are, at any time $t$, of the form: 
\begin{eqnarray*}
\mathrm{Prob}(x < b, t) + \mathrm{Prob}(x > f, t) & = & 1 - \mathrm{Prob}[x \in (b,f), t] \quad \textrm{if} \quad b \le f \,, \\[0.2cm]
\mathrm{Prob}(x < b, t) + \mathrm{Prob}(x > f, t) & = & 1 + \mathrm{Prob}[x \in (b,f), t] \quad \textrm{if} \quad f < b \,.
\end{eqnarray*}
Therefore,
\begin{equation*}
\Delta_{\mathrm{QB}} + \Delta_{\mathrm{QO}} =
\pm \mathrm{Prob}[x \in (b,f), t_2] \mp \mathrm{Prob}[x \in (b,f), t_1] - 1 \le 0 \,,
\end{equation*}
because each probability satisfies  $0 \le \mathrm{Prob}[x \in (b,f), t] \le 1$.

%%%%%%%%%%%%%%%%%%%%%%%%%%%%%%%%%%%%
\subsection{Numerical evaluation of $\sup \Delta$}
%%%%%%%%%%%%%%%%%%%%%%%%%%%%%%%%%%%%

Here, we present details of the numerical estimation of $\sup \{ \lambda \}$ ($= \sup \Delta$) in the eigenproblem
\begin{equation}
	-\frac{1}{\pi} \int_{-\infty}^{+\infty} dv \, (u + v) \sinc(u^2 - v^2) \varphi(v) - \Theta(-u) \varphi(u) = \lambda \, \varphi(u) \,,
	\label{eigprob1}
\end{equation}
where $\sinc(z) \equiv \sin(z) / z$. Truncating the integral, we obtain a new eigenproblem,
\begin{equation*}
	-\frac{1}{\pi} \int_{-L}^{L} dv \, (u + v) \sinc(u^2 - v^2) f(v) - \Theta(-u) f(u) = \lambda_L f(u) \,.
\end{equation*}
Subsequently, discretizing the integral yields yet another eigenproblem,
\begin{equation}
	\sum_{m=-N}^N K_{nm} g_m = \lambda_{L,N} g_n
	\label{eigprob3}
\end{equation}
with
\begin{equation*}
	K_{nm} = -\frac{L^2 (n + m)}{\pi N^2} \sinc \frac{L^2 (n^2 - m^2)}{N^2} - \Theta(-n) \delta_{nm} \,.
\end{equation*}
Our strategy for estimating $\sup \{ \lambda \}$ is based on the assumption that
\begin{equation}
	\sup \{ \lambda \} = \lim_{L \to \infty} \sup \{ \lambda_L \} = \lim_{L \to \infty} \lim_{N \to \infty} \max \{ \lambda_{L,N} \} \,.
	\label{limits}
\end{equation}

We now present values of $\max \{ \lambda_{L,N} \}$ for $L = 10, 15, 20, 25, 30, 35, 40$, obtained using \texttt{linalg.eigh} function from Python's NumPy package. For each $L$, we perform an extrapolation as $N \to \infty$ to evaluate the corresponding value of $\sup \{ \lambda_L \}$. All data in intermediate calculations are presented with full precision (of 16 significant figures), and rounding is applied only to the final result.

\medskip

\noindent \underline{\textbf{Case of $L = 10$}}

\smallskip

Table~\ref{alpha_for_L=10} gives $\max \{ \lambda_{10,N} \}$ for ten different values of $N$.
\begin{table}[h!]
	\centering
	\begin{tabular}{ r||r|l }
		$j$ & $N$ & $\max \{ \lambda_{10,N} \}$ \\
		\hline\hline
		1 & 100 & 0.1089941477648881 \\
		\hline
		2 & 200 & 0.1139918526091255 \\
		\hline
		3 & 300 & 0.1156769692627627 \\
		\hline
		4 & 400 & 0.1165224285063283 \\
		\hline
		5 & 500 & 0.1170305983268448 \\
		\hline
		6 & 600 & 0.1173697450068959 \\
		\hline
		7 & 700 & 0.1176121707518563 \\
		\hline
		8 & 800 & 0.1177940869441254 \\
		\hline
		9 & 900 & 0.1179356345180387 \\
		\hline
		10 & 1000 & 0.1180489085192777
	\end{tabular}
	\caption{The largest eigenvalue in Eq.~\eqref{eigprob3} for $L = 10$ and different $N$.}
	\label{alpha_for_L=10}
\end{table}
We fit $\max \{ \lambda_{10,N} \}$ to a straight line in $1/N$:
\begin{equation*}
	\max \{ \lambda_{10,N} \} = a^{(10)} + \frac{b^{(10)}}{N} \,.
\end{equation*}
The optimal fitting parameters are determined by the linear regression theory (see, e.g., Ref.~\cite{Bevington2003}) and read
\begin{equation*}
	a = \frac{ \left( \sum x_j^2 \right) \left( \sum y_j \right) - \left( \sum x_j \right) \left( \sum x_j y_j \right) }{ J \sum x_j^2 - \left( \sum x_j \right)^2 } \,, \qquad b = \frac{ J \sum x_j y_j - \left( \sum x_j \right) \left( \sum y_j \right) }{ J \sum x_j^2 - \left( \sum x_j \right)^2 } \,,
\end{equation*}
where, for brevity, we have defined $x_j = 1/N_j$, $y_j = \max \{ \lambda_{10,N} \}_j$, and $J = \max \{ j \} = 10$ (see Table~\ref{alpha_for_L=10}). In the present case, these formulas yield $a^{(10)} = 0.1190457811547612$ and $b^{(10)} = -1.006541101888894$.

\begin{figure}[h]
	\centering
	\includegraphics[width=0.6\textwidth]{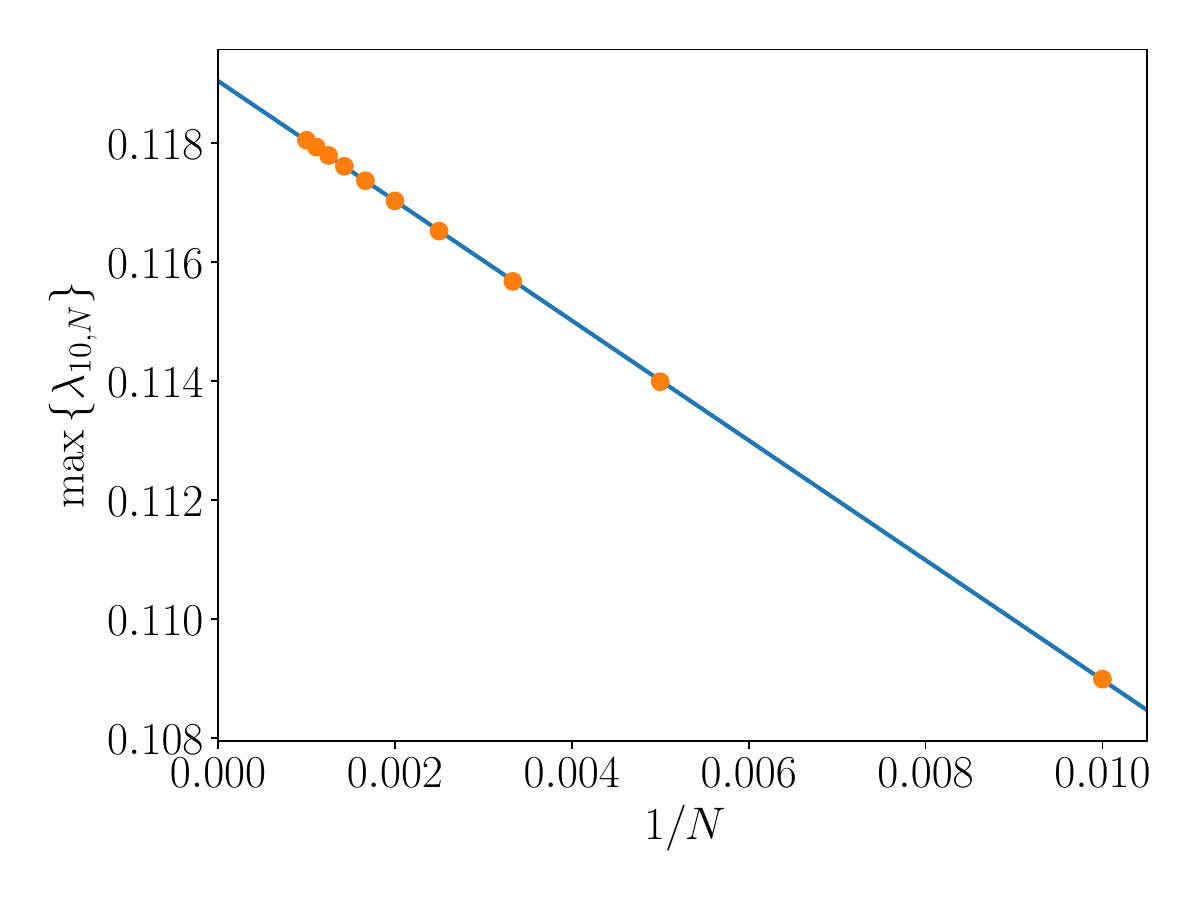}
	\caption{$\max \{ \lambda_{10,N} \}$ versus $1/N$. The orange dots correspond to the values given in Table~\ref{alpha_for_L=10}. The blue line represents the fit detailed in the text.}
	\label{fig:L=10}
\end{figure}
Figure~\ref{fig:L=10} illustrates the accuracy of the fit. The fit residual, defined as $\sum (y_j - a - b x_j)^2$, is $1.1 \times 10^{-9}$.

The uncertainty (or, more precisely, the standard deviation) $\sigma_a$ in the fitting parameter $a$ is given by
\begin{equation*}
	\sigma_a^2 = \frac{\sum (y_j - a - b x_j)^2}{J-2} \cdot \frac{\sum x_j^2}{ J \sum x_j^2 - \left( \sum x_j \right)^2 } \,.
\end{equation*}
In our case, this formula yields $\sigma_{a^{(10)}} = 5.561804879449029 \times 10^{-6}$.
We note that in this and all subsequent calculations, we cross-check our results for the fitting parameters and their associated uncertainties against the output of the \texttt{polyfit} function from Python's NumPy package.

Since $\sup \{ \lambda_{10} \} = \lim_{N \to \infty} \max \{ \lambda_{10,N} \} = a^{(10)}$, with the standard deviation $\sigma_{\sup \{ \lambda_{10} \}} \equiv \sigma_{a^{(10)}}$, we obtain the following estimate:
\begin{equation}
	\sup \{ \lambda_{10} \} = 0.1190457811547612 \,, \qquad \sigma_{\sup \{ \lambda_{10} \}} = 5.561804879449029 \times 10^{-6} \,.
	\label{beta_for_L=10}
\end{equation}

We now repeat the same procedure to obtain the estimates of $\sup \{ \lambda_L \}$ and $\sigma_{\sup \{ \lambda_L \}}$ for $L = 15, 20, 25, 30, 35, 40$.

\medskip

\noindent \underline{\textbf{Case of $L = 15$}}

\smallskip

Table~\ref{alpha_for_L=15} gives $\max \{ \lambda_{15, N} \}$ for ten different values of $N$.
\begin{table}[h!]
	\centering
	\begin{tabular}{ r||r|l }
		$j$ & $N$ & $\max \{ \lambda_{15, N} \}$ \\
		\hline\hline
		1 & 150 & 0.1119434239699512 \\
		\hline
		2 & 300 & 0.1169883703222082 \\
		\hline
		3 & 450 & 0.1186785116904098 \\
		\hline
		4 & 600 & 0.1195264258732301 \\
		\hline
		5 & 750 & 0.1200361158663618 \\
		\hline
		6 & 900 & 0.1203763059276854 \\
		\hline
		7 & 1050 & 0.1206194942121110 \\
		\hline
		8 & 1200 & 0.1208019926220471 \\
		\hline
		9 & 1350 & 0.1209439994908275 \\
		\hline
		10 & 1500 & 0.1210576451546305
	\end{tabular}
	\caption{The largest eigenvalue in Eq.~\eqref{eigprob3} for $L = 15$ and different $N$.}
	\label{alpha_for_L=15}
\end{table}
The fitting procedure yields
\begin{equation*}
	\max \{ \lambda_{15, N} \} = a^{(15)} + \frac{b^{(15)}}{N}
\end{equation*}
with $a^{(15)} = 0.1220638424358060$ and $b^{(15)} = -1.519279315595413$.
\begin{figure}[h]
	\centering
	\includegraphics[width=0.6\textwidth]{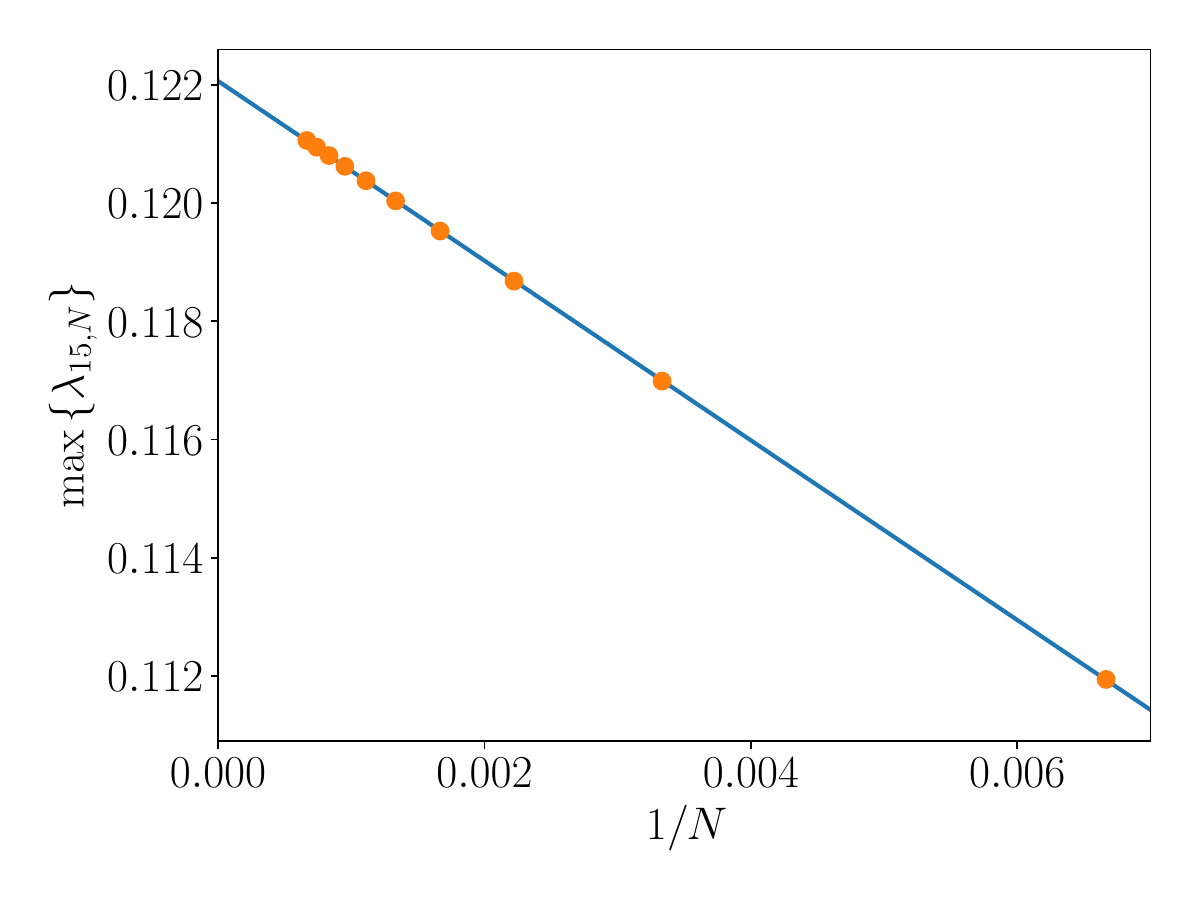}
	\caption{$\max \{ \lambda_{15, N} \}$ versus $1/N$. The orange dots correspond to the values given in Table~\ref{alpha_for_L=15}. The blue line represents the fit detailed in the text.}
	\label{fig:L=15}
\end{figure}
Figure~\ref{fig:L=15} illustrates the accuracy of the fit. The fit residual is $4.1 \times 10^{-10}$.
The standard deviation of $a^{(15)}$ is $\sigma_{a^{(15)}} = 3.375966537451888 \times 10^{-6}$.
Therefore, for $\sup \{ \lambda_{15} \}$ and the corresponding uncertainty, we have
\begin{equation}
	\sup \{ \lambda_{15} \} = 0.1220638424358060 \,, \qquad \sigma_{\sup \{ \lambda_{15} \}} = 3.375966537451888 \times 10^{-6} \,.
	\label{beta_for_L=15}
\end{equation}

\medskip

\noindent \underline{\textbf{Case of $L = 20$}}

\smallskip

Table~\ref{alpha_for_L=20} gives $\max \{ \lambda_{20, N} \}$ for ten different values of $N$.
\begin{table}[h!]
	\centering
	\begin{tabular}{ r||r|l }
		$j$ & $N$ & $\max \{ \lambda_{20, N} \}$ \\
		\hline\hline
		1 & 400 & 0.1184799591245791 \\
		\hline
		2 & 600 & 0.1201715170800806 \\
		\hline
		3 & 800 & 0.1210211077144216 \\
		\hline
		4 & 1000 & 0.1215320463086800 \\
		\hline
		5 & 1200 & 0.1218731553662990 \\
		\hline
		6 & 1400 & 0.1221170385559465 \\
		\hline
		7 & 1600 & 0.1223000778432970 \\
		\hline
		8 & 1800 & 0.1224425165250773 \\
		\hline
		9 & 2000 & 0.1225565144042943 \\
		\hline
		10 & 2200 & 0.1226498163346949
	\end{tabular}
	\caption{The largest eigenvalue in Eq.~\eqref{eigprob3} for $L = 20$ and different $N$.}
	\label{alpha_for_L=20}
\end{table}
The fitting procedure yields
\begin{equation*}
	\max \{ \lambda_{20, N} \} = a^{(20)} + \frac{b^{(20)}}{N}
\end{equation*}
with $a^{(20)} = 0.1235739992757623$ and $b^{(20)} = -2.039355860145372$.
\begin{figure}[h]
	\centering
	\includegraphics[width=0.6\textwidth]{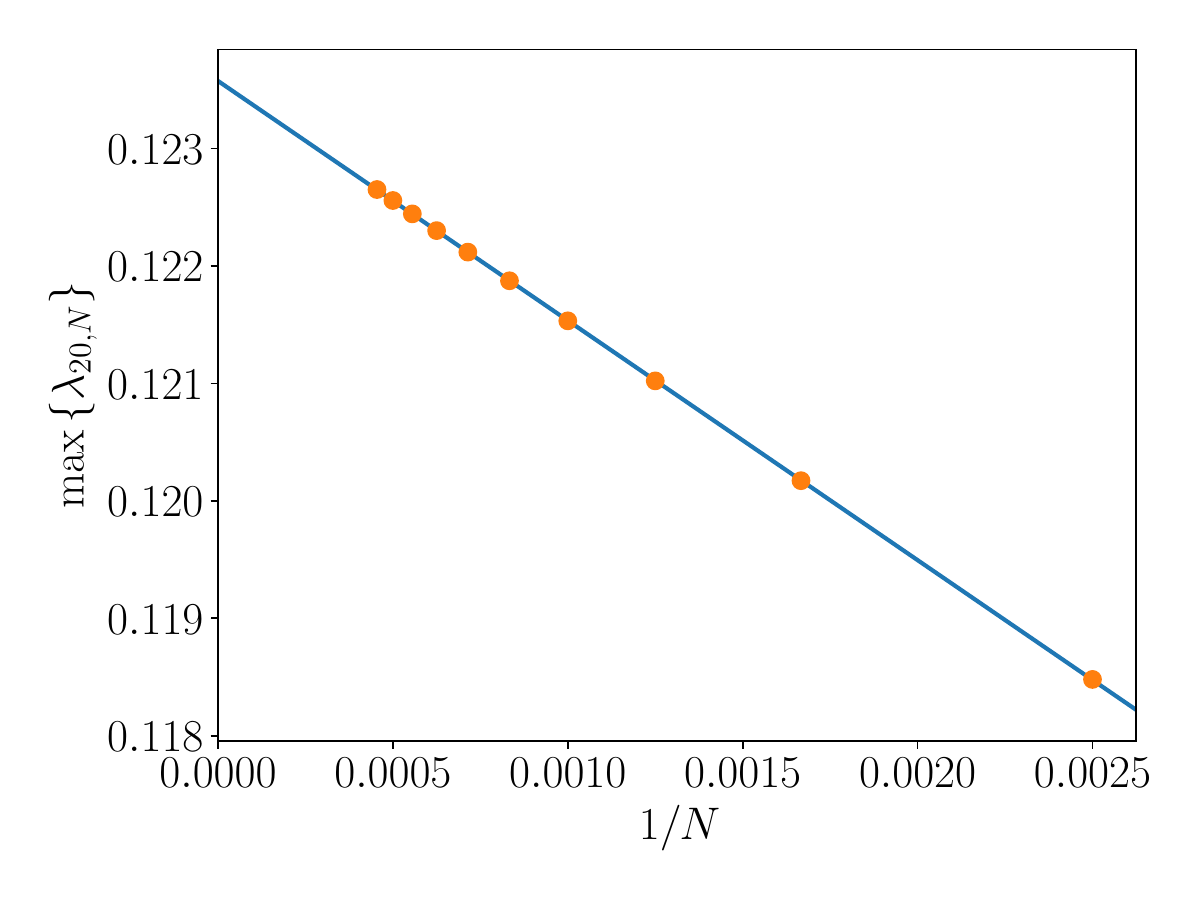}
	\caption{$\max \{ \lambda_{20, N} \}$ versus $1/N$. The orange dots correspond to the values given in Table~\ref{alpha_for_L=20}. The blue line represents the fit detailed in the text.}
	\label{fig:L=20}
\end{figure}
Figure~\ref{fig:L=20} illustrates the accuracy of the fit. The fit residual is $6.9 \times 10^{-11}$.
The standard deviation of $a^{(20)}$ is $\sigma_{a^{(20)}} = 1.794581838275372 \times 10^{-6}$.
Therefore, for $\sup \{ \lambda_{20} \}$ and the corresponding uncertainty, we have
\begin{equation}
	\sup \{ \lambda_{20} \} = 0.1235739992757623 \,, \qquad \sigma_{\sup \{ \lambda_{20} \}} = 1.794581838275372 \times 10^{-6} \,.
	\label{beta_for_L=20}
\end{equation}

\medskip

\noindent \underline{\textbf{Case of $L = 25$}}

\smallskip

Table~\ref{alpha_for_L=25} gives $\max \{ \lambda_{25, N} \}$ for ten different values of $N$.
\begin{table}[h!]
	\centering
	\begin{tabular}{ r||r|l }
		$j$ & $N$ & $\max \{ \lambda_{25, N} \}$ \\
		\hline\hline
		1 & 500 & 0.1193720192594545 \\
		\hline
		2 & 750 & 0.1210700920114530 \\
		\hline
		3 & 1000 & 0.1219223352025476 \\
		\hline
		4 & 1250 & 0.1224346215338796 \\
		\hline
		5 & 1500 & 0.1227765382427701 \\
		\hline
		6 & 1750 & 0.1230209570947946 \\
		\hline
		7 & 2000 & 0.1232043766828351 \\
		\hline
		8 & 2250 & 0.1233470988716490 \\
		\hline
		9 & 2500 & 0.1234613160028462 \\
		\hline
		10 & 2750 & 0.1235547924141025
	\end{tabular}
	\caption{The largest eigenvalue in Eq.~\eqref{eigprob3} for $L = 25$ and different $N$.}
	\label{alpha_for_L=25}
\end{table}
The fitting procedure yields
\begin{equation*}
	\max \{ \lambda_{25, N} \} = a^{(25)} + \frac{b^{(25)}}{N}
\end{equation*}
with $a^{(25)} = 0.1244822862467697$ and $b^{(25)} = -2.556926934667362$.
\begin{figure}[h]
	\centering
	\includegraphics[width=0.6\textwidth]{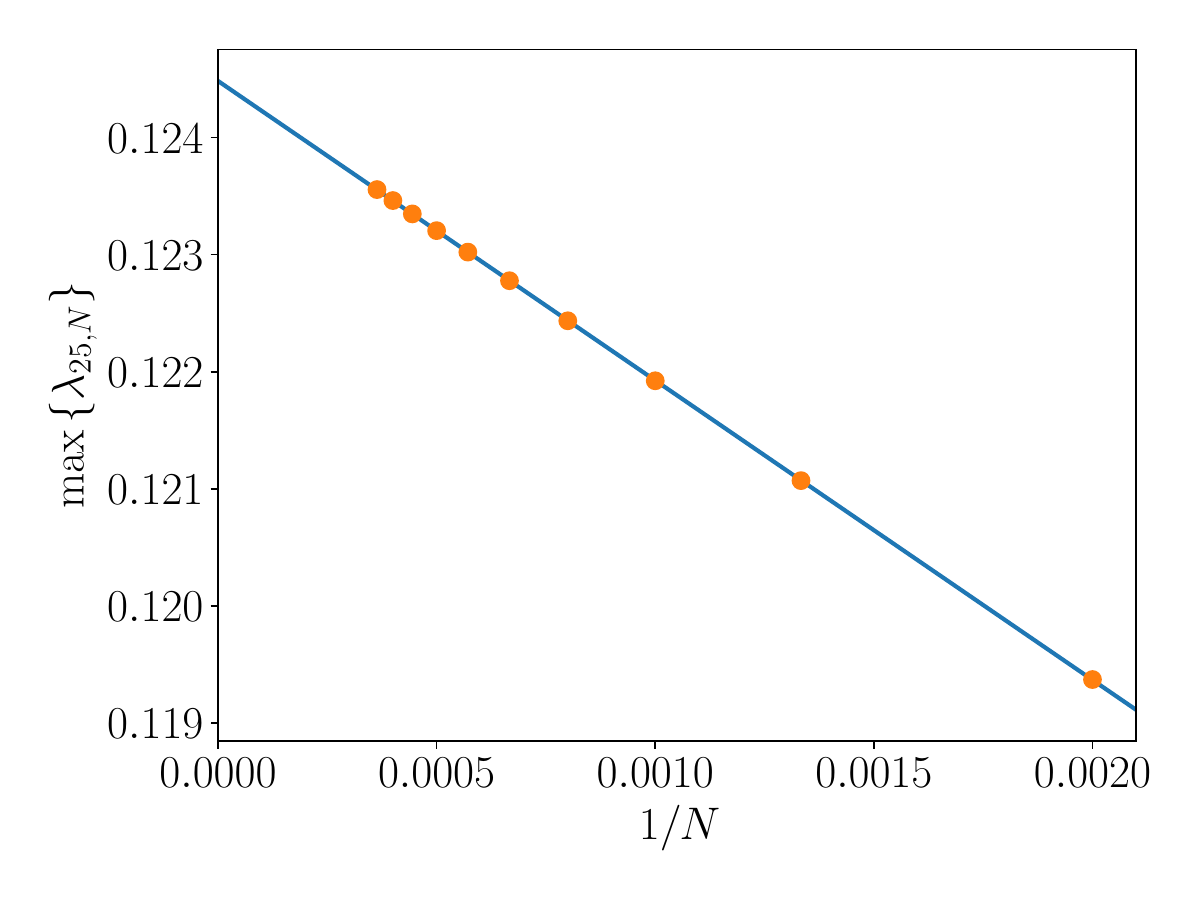}
	\caption{$\max \{ \lambda_{25, N} \}$ versus $1/N$. The orange dots correspond to the values given in Table~\ref{alpha_for_L=25}. The blue line represents the fit detailed in the text.}
	\label{fig:L=25}
\end{figure}
Figure~\ref{fig:L=25} illustrates the accuracy of the fit. The fit residual is $4.7 \times 10^{-11}$.
The standard deviation of $a^{(25)}$ is $\sigma_{a^{(25)}} = 1.477042007656790 \times 10^{-6}$.
Therefore, for $\sup \{ \lambda_{25} \}$ and the corresponding uncertainty, we have
\begin{equation}
	\sup \{ \lambda_{25} \} = 0.1244822862467697 \,, \qquad \sigma_{\sup \{ \lambda_{25} \}} = 1.477042007656790 \times 10^{-6} \,.
	\label{beta_for_L=25}
\end{equation}

\medskip

\noindent \underline{\textbf{Case of $L = 30$}}

\smallskip

Table~\ref{alpha_for_L=30} gives $\max \{ \lambda_{30, N} \}$ for ten different values of $N$.
\begin{table}[h!]
	\centering
	\begin{tabular}{ r||r|l }
		$j$ & $N$ & $\max \{ \lambda_{30, N} \}$ \\
		\hline\hline
		1 & 600 & 0.1199705773485895 \\
		\hline
		2 & 900 & 0.1216727154735607 \\
		\hline
		3 & 1200 & 0.1225237102038637 \\
		\hline
		4 & 1500 & 0.1230354498337929 \\
		\hline
		5 & 1800 & 0.1233770750757993 \\
		\hline
		6 & 2100 & 0.1236213172702387 \\
		\hline
		7 & 2400 & 0.1238046200925763 \\
		\hline
		8 & 2700 & 0.1239472601723789 \\
		\hline
		9 & 3000 & 0.1240614168298270 \\
		\hline
		10 & 3300 & 0.1241548470766964
	\end{tabular}
	\caption{The largest eigenvalue in Eq.~\eqref{eigprob3} for $L = 30$ and different $N$.}
	\label{alpha_for_L=30}
\end{table}
The fitting procedure yields
\begin{equation*}
	\max \{ \lambda_{30, N} \} = a^{(30)} + \frac{b^{(30)}}{N}
\end{equation*}
with $a^{(30)} = 0.1250829195703286$ and $b^{(30)} = -3.068533796622797$.
\begin{figure}[h]
	\centering
	\includegraphics[width=0.6\textwidth]{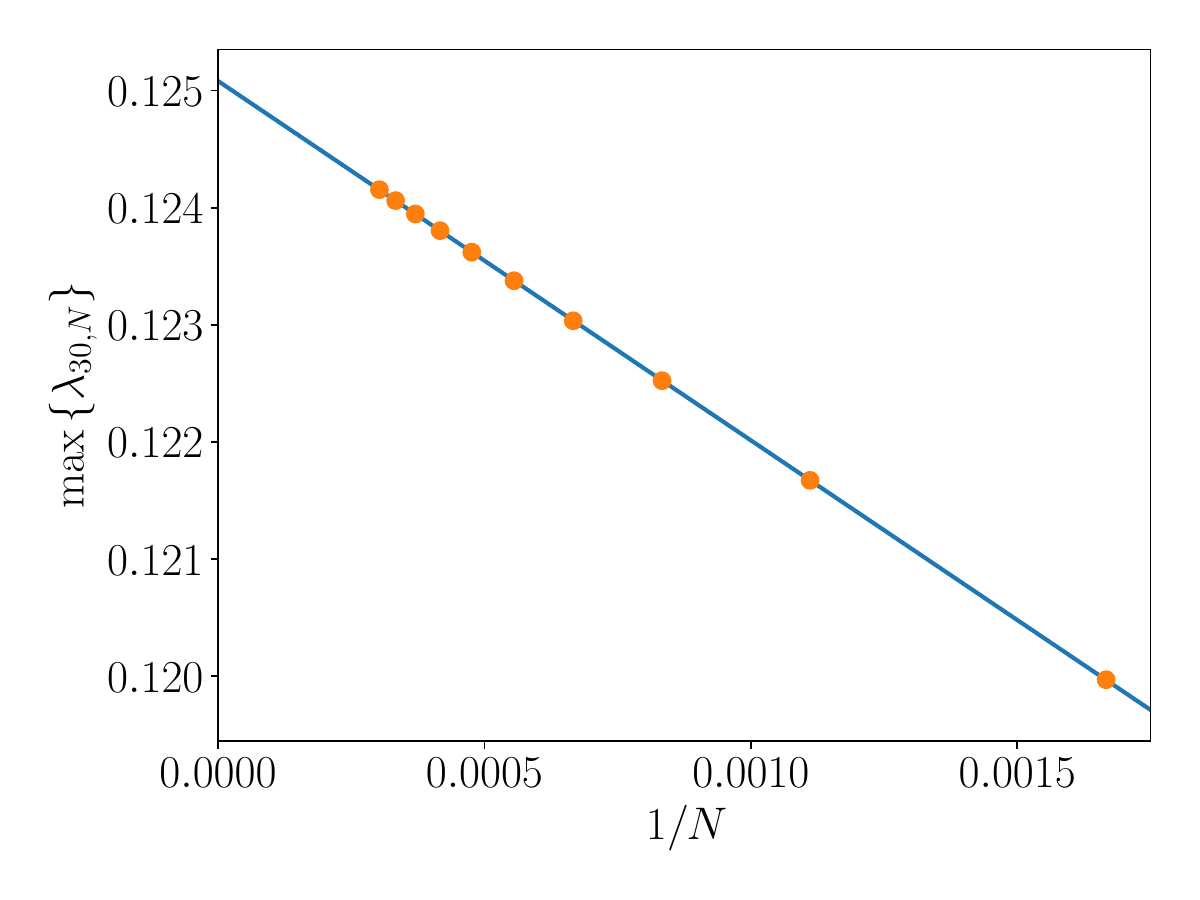}
	\caption{$\max \{ \lambda_{30, N} \}$ versus $1/N$. The orange dots correspond to the values given in Table~\ref{alpha_for_L=30}. The blue line represents the fit detailed in the text.}
	\label{fig:L=30}
\end{figure}
Figure~\ref{fig:L=30} illustrates the accuracy of the fit. The fit residual is $1.9 \times 10^{-11}$.
The standard deviation of $a^{(30)}$ is $\sigma_{a^{(30)}} = 9.338351587060546 \times 10^{-7}$.
Therefore, for $\sup \{ \lambda_{30} \}$ and the corresponding uncertainty, we have
\begin{equation}
	\sup \{ \lambda_{30} \} = 0.1250829195703286 \,, \qquad \sigma_{\sup \{ \lambda_{30} \}} = 9.338351587060546 \times 10^{-7} \,.
	\label{beta_for_L=30}
\end{equation}

\medskip

\noindent \underline{\textbf{Case of $L = 35$}}

\smallskip

Table~\ref{alpha_for_L=35} gives $\max \{ \lambda_{35, N} \}$ for ten different values of $N$.
\begin{table}[h!]
	\centering
	\begin{tabular}{ r||r|l }
		$j$ & $N$ & $\max \{ \lambda_{35, N} \}$ \\
		\hline\hline
		1 & 800 & 0.1210457163111231 \\
		\hline
		2 & 1200 & 0.1225236257266571 \\
		\hline
		3 & 1600 & 0.1232707319394567 \\
		\hline
		4 & 2000 & 0.1237196617010170 \\
		\hline
		5 & 2400 & 0.1240192405176445 \\
		\hline
		6 & 2800 & 0.1242333716554072 \\
		\hline
		7 & 3200 & 0.1243940508807573 \\
		\hline
		8 & 3600 & 0.1245190718418926 \\
		\hline
		9 & 4000 & 0.1246191191181317 \\
		\hline
		10 & 4400 & 0.1247009962047494
	\end{tabular}
	\caption{The largest eigenvalue in Eq.~\eqref{eigprob3} for $L = 35$ and different $N$.}
	\label{alpha_for_L=35}
\end{table}
The fitting procedure yields
\begin{equation*}
	\max \{ \lambda_{35, N} \} = a^{(35)} + \frac{b^{(35)}}{N}
\end{equation*}
with $a^{(35)} = 0.1255108771862945$ and $b^{(35)} = -3.577085710064331$.
\begin{figure}[h]
	\centering
	\includegraphics[width=0.6\textwidth]{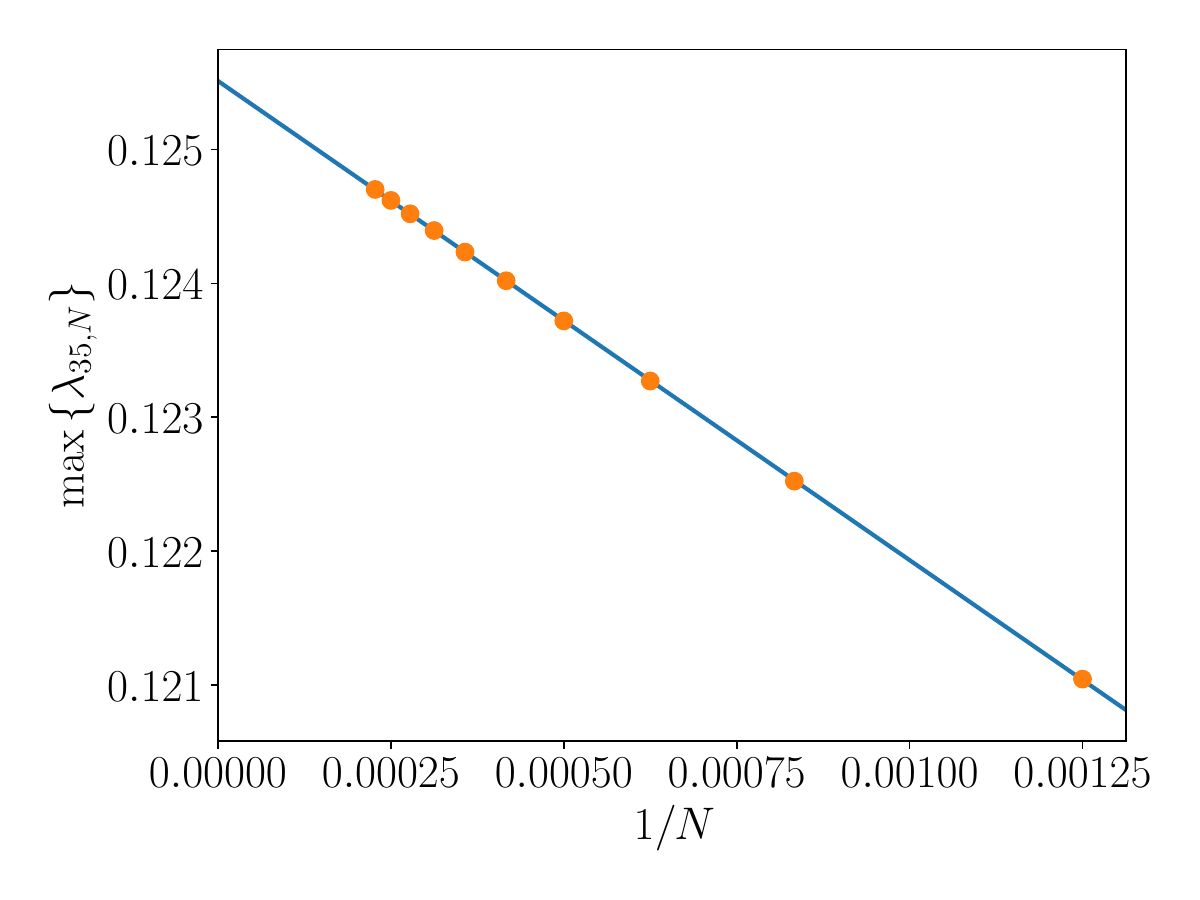}
	\caption{$\max \{ \lambda_{35, N} \}$ versus $1/N$. The orange dots correspond to the values given in Table~\ref{alpha_for_L=35}. The blue line represents the fit detailed in the text.}
	\label{fig:L=35}
\end{figure}
Figure~\ref{fig:L=35} illustrates the accuracy of the fit. The fit residual is $1.3 \times 10^{-10}$.
The standard deviation of $a^{(35)}$ is $\sigma_{a^{(35)}} = 2.433955824900637 \times 10^{-6}$.
Therefore, for $\sup \{ \lambda_{35} \}$ and the corresponding uncertainty, we have
\begin{equation}
	\sup \{ \lambda_{35} \} = 0.1255108771862945 \,, \qquad \sigma_{\sup \{ \lambda_{35} \}} = 2.433955824900637 \times 10^{-6} \,.
	\label{beta_for_L=35}
\end{equation}

\medskip

\noindent \underline{\textbf{Case of $L = 40$}}

\smallskip

Table~\ref{alpha_for_L=40} gives $\max \{ \lambda_{40, N} \}$ for eight different values of $N$.
\begin{table}[h!]
	\centering
	\begin{tabular}{ r||r|l }
		$j$ & $N$ & $\max \{ \lambda_{40, N} \}$ \\
		\hline\hline
		1 & 1200 & 0.1224211080034019 \\
		\hline
		2 & 1800 & 0.1235567482529446 \\
		\hline
		3 & 2400 & 0.1241265265788500 \\
		\hline
		4 & 3000 & 0.1244689068130608 \\
		\hline
		5 & 3600 & 0.1246973635442337 \\
		\hline
		6 & 4200 & 0.1248606442986848 \\
		\hline
		7 & 4800 & 0.1249831574822948 \\
		\hline
		8 & 5400 & 0.1250784764658047
	\end{tabular}
	\caption{The largest eigenvalue in Eq.~\eqref{eigprob3} for $L = 40$ and different $N$.}
	\label{alpha_for_L=40}
\end{table}
The fitting procedure yields
\begin{equation*}
	\max \{ \lambda_{40, N} \} = a^{(40)} + \frac{b^{(40)}}{N}
\end{equation*}
with $a^{(40)} = 0.1258365731314665$ and $b^{(40)} = -4.100558963340225$.
\begin{figure}[h]
	\centering
	\includegraphics[width=0.6\textwidth]{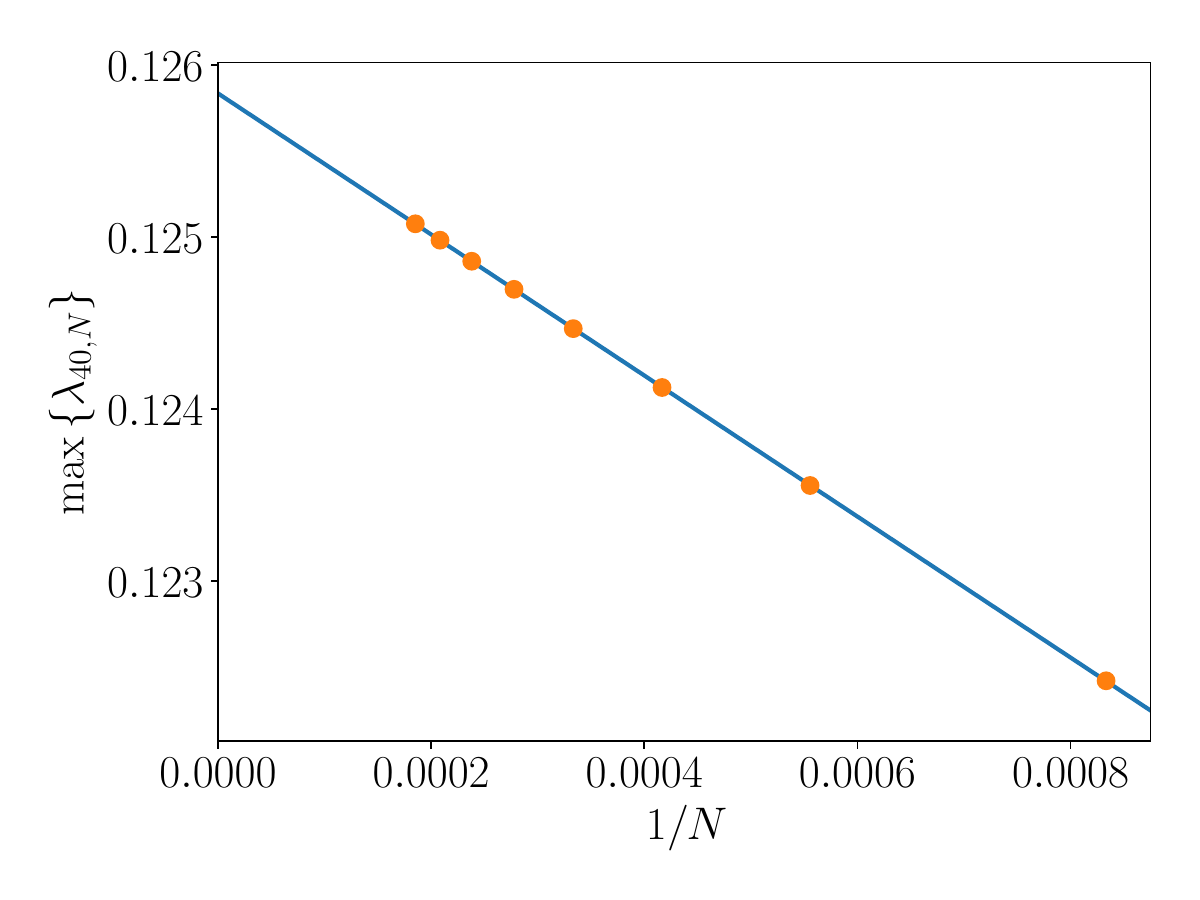}
	\caption{$\max \{ \lambda_{40, N} \}$ versus $1/N$. The orange dots correspond to the values given in Table~\ref{alpha_for_L=40}. The blue line represents the fit detailed in the text.}
	\label{fig:L=40}
\end{figure}
Figure~\ref{fig:L=40} illustrates the accuracy of the fit. The fit residual is $1.1 \times 10^{-11}$.
The standard deviation of $a^{(40)}$ is $\sigma_{a^{(40)}} = 1.017018984296298 \times 10^{-6}$.
Therefore, for $\sup \{ \lambda_{40} \}$ and the corresponding uncertainty, we have
\begin{equation}
	\sup \{ \lambda_{40} \} = 0.1258365731314665 \,, \qquad \sigma_{\sup \{ \lambda_{40} \}} = 1.017018984296298 \times 10^{-6} \,.
	\label{beta_for_L=40}
\end{equation}

\medskip

\noindent \underline{\textbf{Limit of $\sup \{ \lambda_L \}$ as $L \to \infty$}}

\smallskip

Table~\ref{beta_vs_L} summarizes the values for $\sup \{ \lambda_L \}$ and its standard deviation $\sigma_{\sup \{ \lambda_L \}}$ obtained in Eqs.~(\ref{beta_for_L=10}--\ref{beta_for_L=40}).
\begin{table}[h!]
	\centering
	\begin{tabular}{ r||r|l|l }
		$j$ & $L$ & $\sup \{ \lambda_L \}$ & $\sigma_{\sup \{ \lambda_L \}}$ \\
		\hline\hline
		1 & 10 & 0.1190457811547612 & $5.561804879449029 \times 10^{-6}$ \\
		\hline
		2 & 15 & 0.1220638424358060 & $3.375966537451888 \times 10^{-6}$ \\
		\hline
		3 & 20 & 0.1235739992757623 & $1.794581838275372 \times 10^{-6}$ \\
		\hline
		4 & 25 & 0.1244822862467697 & $1.477042007656790 \times 10^{-6}$ \\
		\hline
		5 & 30 & 0.1250829195703286 & $9.338351587060546 \times 10^{-7}$ \\
		\hline
		6 & 35 & 0.1255108771862945 & $2.433955824900637 \times 10^{-6}$ \\
		\hline
		7 & 40 & 0.1258365731314665 & $1.017018984296298 \times 10^{-6}$
	\end{tabular}
	\caption{The largest eigenvalues in Eq.~\eqref{eigprob3}, along with their uncertainties, for different values of $L$.}
	\label{beta_vs_L}
\end{table}
We now fit $\sup \{ \lambda_L \}$ to a straight line in $1/L$:
\begin{equation*}
	\sup \{ \lambda_L \} = a + \frac{b}{L} \,.
\end{equation*}
The optimal fitting parameters are given by
\begin{equation*}
	a = \frac{ \left( \sum \frac{x_j^2}{\sigma_j^2} \right) \left( \sum \frac{y_j}{\sigma_j^2} \right) - \left( \sum \frac{x_j}{\sigma_j^2} \right) \left( \sum \frac{x_j y_j}{\sigma_j^2} \right) }{ \left( \sum \frac{1}{\sigma_j^2} \right) \left( \sum \frac{x_j^2}{\sigma_j^2} \right) - \left( \sum \frac{x_j}{\sigma_j^2} \right)^2 } \,, \qquad b = \frac{ \left( \sum \frac{1}{\sigma_j^2} \right) \left( \sum \frac{x_j y_j}{\sigma_j^2} \right) - \left( \sum \frac{x_j}{\sigma_j^2} \right) \left( \sum \frac{y_j}{\sigma_j^2} \right) }{ \left( \sum \frac{1}{\sigma_j^2} \right) \left( \sum \frac{x_j^2}{\sigma_j^2} \right) - \left( \sum \frac{x_j}{\sigma_j^2} \right)^2 } \,,
\end{equation*}
where, for brevity, $x_j = L_j$, $y_j = \sup \{ \lambda_L \}_j$, and $\sigma_j = \left( \sigma_{\sup \{ \lambda_L \}} \right)_j$ (see Table~\ref{beta_vs_L}). In the present case, these formulas yield
\begin{equation*}
	a = 0.1280997589328653 \,, \qquad b = -9.050752023369246 \times 10^{-2} \,.
\end{equation*}

\begin{figure}[h]
	\centering
	\includegraphics[width=0.6\textwidth]{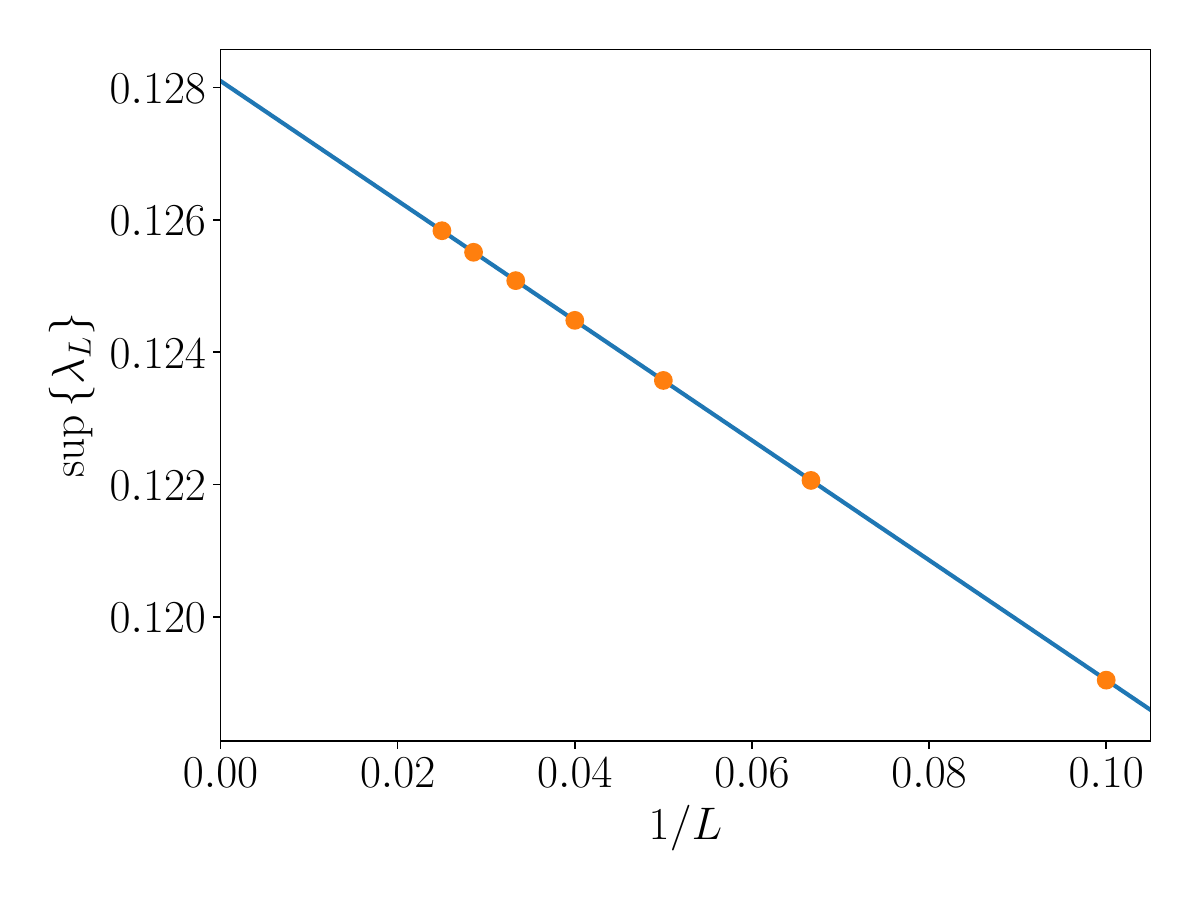}
	\caption{$\sup \{ \lambda_L \}$ versus $1/L$. The orange dots correspond to the values given in Table~\ref{beta_vs_L}. The blue line represents the fit detailed in the text.}
	\label{fig:beta}
\end{figure}
Figure~\ref{fig:beta} illustrates the accuracy of the fit. The fit residual is $3.2 \times 10^{-11}$.

The standard deviation $\sigma_a$ of the fitting parameter $a$ is given by
\begin{equation*}
	\sigma_a^2 = \frac{\sum \frac{(y_j - a - b x_j)^2}{\sigma_j^2}}{J-2} \cdot \frac{ \sum \frac{x_j^2}{\sigma_j^2} }{ \left( \sum \frac{1}{\sigma_j^2} \right) \left( \sum \frac{x_j^2}{\sigma_j^2} \right) - \left( \sum \frac{x_j}{\sigma_j^2} \right)^2 } \,,
\end{equation*}
which, in the present case, gives
\begin{equation*}
	\sigma_a = 1.995873910685532 \times 10^{-6} \,.
\end{equation*}

Finally, in view of Eq.~\eqref{limits}, we state our estimate for the supremum of the eigenvalue spectrum in Eq.~\eqref{eigprob1}:
\begin{equation*}
	\sup \{ \lambda \} = 0.128100 \pm 0.000002 \,.
\end{equation*}
As a reminder, the quoted uncertainty corresponds to one standard deviation, representing a 68\% confidence level.

\end{widetext}


\begin{thebibliography}{34}%
\makeatletter
\providecommand \@ifxundefined [1]{%
 \@ifx{#1\undefined}
}%
\providecommand \@ifnum [1]{%
 \ifnum #1\expandafter \@firstoftwo
 \else \expandafter \@secondoftwo
 \fi
}%
\providecommand \@ifx [1]{%
 \ifx #1\expandafter \@firstoftwo
 \else \expandafter \@secondoftwo
 \fi
}%
\providecommand \natexlab [1]{#1}%
\providecommand \enquote  [1]{``#1''}%
\providecommand \bibnamefont  [1]{#1}%
\providecommand \bibfnamefont [1]{#1}%
\providecommand \citenamefont [1]{#1}%
\providecommand \href@noop [0]{\@secondoftwo}%
\providecommand \href [0]{\begingroup \@sanitize@url \@href}%
\providecommand \@href[1]{\@@startlink{#1}\@@href}%
\providecommand \@@href[1]{\endgroup#1\@@endlink}%
\providecommand \@sanitize@url [0]{\catcode `\\12\catcode `\$12\catcode
  `\&12\catcode `\#12\catcode `\^12\catcode `\_12\catcode `\%12\relax}%
\providecommand \@@startlink[1]{}%
\providecommand \@@endlink[0]{}%
\providecommand \url  [0]{\begingroup\@sanitize@url \@url }%
\providecommand \@url [1]{\endgroup\@href {#1}{\urlprefix }}%
\providecommand \urlprefix  [0]{URL }%
\providecommand \Eprint [0]{\href }%
\providecommand \doibase [0]{https://doi.org/}%
\providecommand \selectlanguage [0]{\@gobble}%
\providecommand \bibinfo  [0]{\@secondoftwo}%
\providecommand \bibfield  [0]{\@secondoftwo}%
\providecommand \translation [1]{[#1]}%
\providecommand \BibitemOpen [0]{}%
\providecommand \bibitemStop [0]{}%
\providecommand \bibitemNoStop [0]{.\EOS\space}%
\providecommand \EOS [0]{\spacefactor3000\relax}%
\providecommand \BibitemShut  [1]{\csname bibitem#1\endcsname}%
\let\auto@bib@innerbib\@empty
%</preamble>
\bibitem [{\citenamefont {Allcock}(1969)}]{Allcock1969}%
  \BibitemOpen
  \bibfield  {author} {\bibinfo {author} {\bibfnamefont {G.}~\bibnamefont
  {Allcock}},\ }\bibfield  {title} {\bibinfo {title} {{The time of arrival in
  quantum mechanics III. The measurement ensemble}},\ }\href
  {https://doi.org/10.1016/0003-4916(69)90253-x} {\bibfield  {journal}
  {\bibinfo  {journal} {Ann. Phys.}\ }\textbf {\bibinfo {volume} {53}},\
  \bibinfo {pages} {311–348} (\bibinfo {year} {1969})}\BibitemShut {NoStop}%
\bibitem [{\citenamefont {Kijowski}(1974)}]{Kijowski1974}%
  \BibitemOpen
  \bibfield  {author} {\bibinfo {author} {\bibfnamefont {J.}~\bibnamefont
  {Kijowski}},\ }\bibfield  {title} {\bibinfo {title} {{On the time operator in
  quantum mechanics and the Heisenberg uncertainty relation for energy and
  time}},\ }\href {https://doi.org/10.1016/s0034-4877(74)80004-2} {\bibfield
  {journal} {\bibinfo  {journal} {Rep. Math. Phys.}\ }\textbf {\bibinfo
  {volume} {6}},\ \bibinfo {pages} {361–386} (\bibinfo {year}
  {1974})}\BibitemShut {NoStop}%
\bibitem [{\citenamefont {Werner}(1988)}]{Wer88Wigner}%
  \BibitemOpen
  \bibfield  {author} {\bibinfo {author} {\bibfnamefont {R.~F.}\ \bibnamefont
  {Werner}},\ }\bibfield  {title} {\bibinfo {title} {{Wigner quantisation of
  arrival time and oscillator phase}},\ }\href
  {https://doi.org/10.1088/0305-4470/21/24/012} {\bibfield  {journal} {\bibinfo
   {journal} {J. Phys. A: Math. Gen.}\ }\textbf {\bibinfo {volume} {21}},\
  \bibinfo {pages} {4565} (\bibinfo {year} {1988})}\BibitemShut {NoStop}%
\bibitem [{\citenamefont {Bracken}\ and\ \citenamefont
  {Melloy}(1994)}]{Bracken1994}%
  \BibitemOpen
  \bibfield  {author} {\bibinfo {author} {\bibfnamefont {A.~J.}\ \bibnamefont
  {Bracken}}\ and\ \bibinfo {author} {\bibfnamefont {G.~F.}\ \bibnamefont
  {Melloy}},\ }\bibfield  {title} {\bibinfo {title} {{Probability backflow and
  a new dimensionless quantum number}},\ }\href
  {https://doi.org/10.1088/0305-4470/27/6/040} {\bibfield  {journal} {\bibinfo
  {journal} {J. Phys, A: Math. Gen.}\ }\textbf {\bibinfo {volume} {27}},\
  \bibinfo {pages} {2197–2211} (\bibinfo {year} {1994})}\BibitemShut
  {NoStop}%
\bibitem [{\citenamefont {Bracken}(2021)}]{Bracken2021}%
  \BibitemOpen
  \bibfield  {author} {\bibinfo {author} {\bibfnamefont {A.~J.}\ \bibnamefont
  {Bracken}},\ }\bibfield  {title} {\bibinfo {title} {{Probability flow for a
  free particle: new quantum effects}},\ }\href
  {https://doi.org/10.1088/1402-4896/abdd54} {\bibfield  {journal} {\bibinfo
  {journal} {Phys. Scr.}\ }\textbf {\bibinfo {volume} {96}},\ \bibinfo {pages}
  {045201} (\bibinfo {year} {2021})}\BibitemShut {NoStop}%
\bibitem [{\citenamefont {Goussev}(2019)}]{Goussev2019}%
  \BibitemOpen
  \bibfield  {author} {\bibinfo {author} {\bibfnamefont {A.}~\bibnamefont
  {Goussev}},\ }\bibfield  {title} {\bibinfo {title} {{Equivalence between
  quantum backflow and classically forbidden probability flow in a
  diffraction-in-time problem}},\ }\href
  {https://doi.org/10.1103/physreva.99.043626} {\bibfield  {journal} {\bibinfo
  {journal} {Phys. Rev. A}\ }\textbf {\bibinfo {volume} {99}},\ \bibinfo
  {pages} {043626} (\bibinfo {year} {2019})}\BibitemShut {NoStop}%
\bibitem [{\citenamefont {Tsirelson}()}]{Tsirelson2006}%
  \BibitemOpen
  \bibfield  {author} {\bibinfo {author} {\bibfnamefont {B.}~\bibnamefont
  {Tsirelson}},\ }\bibfield  {title} {\bibinfo {title} {{How often is the
  coordinate of a harmonic oscillator positive?}},\ }\href@noop {} {\ }\bibinfo
  {note} {{arXiv:2505.13184}}\BibitemShut {NoStop}%
\bibitem [{\citenamefont {Zaw}\ \emph {et~al.}(2022)\citenamefont {Zaw},
  \citenamefont {Aw}, \citenamefont {Lasmar},\ and\ \citenamefont
  {Scarani}}]{Zaw2022}%
  \BibitemOpen
  \bibfield  {author} {\bibinfo {author} {\bibfnamefont {L.~H.}\ \bibnamefont
  {Zaw}}, \bibinfo {author} {\bibfnamefont {C.~C.}\ \bibnamefont {Aw}},
  \bibinfo {author} {\bibfnamefont {Z.}~\bibnamefont {Lasmar}},\ and\ \bibinfo
  {author} {\bibfnamefont {V.}~\bibnamefont {Scarani}},\ }\bibfield  {title}
  {\bibinfo {title} {{Detecting quantumness in uniform precessions}},\ }\href
  {https://doi.org/10.1103/physreva.106.032222} {\bibfield  {journal} {\bibinfo
   {journal} {Phys. Rev. A}\ }\textbf {\bibinfo {volume} {106}},\ \bibinfo
  {pages} {032222} (\bibinfo {year} {2022})}\BibitemShut {NoStop}%
\bibitem [{\citenamefont {Zaw}\ \emph {et~al.}(2025)\citenamefont {Zaw},
  \citenamefont {Weilenmann},\ and\ \citenamefont {Scarani}}]{Zaw2025prl}%
  \BibitemOpen
  \bibfield  {author} {\bibinfo {author} {\bibfnamefont {L.~H.}\ \bibnamefont
  {Zaw}}, \bibinfo {author} {\bibfnamefont {M.}~\bibnamefont {Weilenmann}},\
  and\ \bibinfo {author} {\bibfnamefont {V.}~\bibnamefont {Scarani}},\
  }\bibfield  {title} {\bibinfo {title} {{Tsirelson’s Inequality for the
  Precession Protocol Is Maximally Violated by Quantum Theory}},\ }\href
  {https://doi.org/10.1103/physrevlett.134.190201} {\bibfield  {journal}
  {\bibinfo  {journal} {Phys. Rev. Lett.}\ }\textbf {\bibinfo {volume} {134}},\
  \bibinfo {pages} {190201} (\bibinfo {year} {2025})}\BibitemShut {NoStop}%
\bibitem [{\citenamefont {Zaw}\ and\ \citenamefont {Scarani}(2025)}]{Zaw2025}%
  \BibitemOpen
  \bibfield  {author} {\bibinfo {author} {\bibfnamefont {L.~H.}\ \bibnamefont
  {Zaw}}\ and\ \bibinfo {author} {\bibfnamefont {V.}~\bibnamefont {Scarani}},\
  }\bibfield  {title} {\bibinfo {title} {{All three-angle variants of
  Tsirelson’s precession protocol, and improved bounds for wedge integrals of
  Wigner functions}},\ }\href {https://doi.org/10.1038/s41534-025-01098-7}
  {\bibfield  {journal} {\bibinfo  {journal} {npj Quan. Inf.}\ }\textbf
  {\bibinfo {volume} {11}},\ \bibinfo {pages} {152} (\bibinfo {year}
  {2025})}\BibitemShut {NoStop}%
\bibitem [{\citenamefont {Vaartjes}\ \emph {et~al.}(2025)\citenamefont
  {Vaartjes}, \citenamefont {Nurizzo}, \citenamefont {Zaw}, \citenamefont
  {Wilhelm}, \citenamefont {Yu}, \citenamefont {Holmes}, \citenamefont
  {Schwienbacher}, \citenamefont {Kringhøj}, \citenamefont {van Blankenstein},
  \citenamefont {Jakob}, \citenamefont {Hudson}, \citenamefont {Itoh},
  \citenamefont {Murray}, \citenamefont {Blume-Kohout}, \citenamefont {Anand},
  \citenamefont {Dzurak}, \citenamefont {Jamieson}, \citenamefont {Scarani},\
  and\ \citenamefont {Morello}}]{Vaartjes2025}%
  \BibitemOpen
  \bibfield  {author} {\bibinfo {author} {\bibfnamefont {A.}~\bibnamefont
  {Vaartjes}}, \bibinfo {author} {\bibfnamefont {M.}~\bibnamefont {Nurizzo}},
  \bibinfo {author} {\bibfnamefont {L.~H.}\ \bibnamefont {Zaw}}, \bibinfo
  {author} {\bibfnamefont {B.}~\bibnamefont {Wilhelm}}, \bibinfo {author}
  {\bibfnamefont {X.}~\bibnamefont {Yu}}, \bibinfo {author} {\bibfnamefont
  {D.}~\bibnamefont {Holmes}}, \bibinfo {author} {\bibfnamefont
  {D.}~\bibnamefont {Schwienbacher}}, \bibinfo {author} {\bibfnamefont
  {A.}~\bibnamefont {Kringhøj}}, \bibinfo {author} {\bibfnamefont {M.~R.}\
  \bibnamefont {van Blankenstein}}, \bibinfo {author} {\bibfnamefont {A.~M.}\
  \bibnamefont {Jakob}}, \bibinfo {author} {\bibfnamefont {F.~E.}\ \bibnamefont
  {Hudson}}, \bibinfo {author} {\bibfnamefont {K.~M.}\ \bibnamefont {Itoh}},
  \bibinfo {author} {\bibfnamefont {R.~J.}\ \bibnamefont {Murray}}, \bibinfo
  {author} {\bibfnamefont {R.}~\bibnamefont {Blume-Kohout}}, \bibinfo {author}
  {\bibfnamefont {N.}~\bibnamefont {Anand}}, \bibinfo {author} {\bibfnamefont
  {A.~S.}\ \bibnamefont {Dzurak}}, \bibinfo {author} {\bibfnamefont {D.~N.}\
  \bibnamefont {Jamieson}}, \bibinfo {author} {\bibfnamefont {V.}~\bibnamefont
  {Scarani}},\ and\ \bibinfo {author} {\bibfnamefont {A.}~\bibnamefont
  {Morello}},\ }\bibfield  {title} {\bibinfo {title} {{Certifying the
  quantumness of a nuclear spin qudit through its uniform precession}},\ }\href
  {https://doi.org/10.1016/j.newton.2025.100017} {\bibfield  {journal}
  {\bibinfo  {journal} {Newton}\ }\textbf {\bibinfo {volume} {1}},\ \bibinfo
  {pages} {100017} (\bibinfo {year} {2025})}\BibitemShut {NoStop}%
\bibitem [{\citenamefont {Garg}\ \emph {et~al.}()\citenamefont {Garg},
  \citenamefont {Halliwell},\ and\ \citenamefont
  {Venkataraman}}]{Garg2025_arXiv}%
  \BibitemOpen
  \bibfield  {author} {\bibinfo {author} {\bibfnamefont {A.}~\bibnamefont
  {Garg}}, \bibinfo {author} {\bibfnamefont {J.}~\bibnamefont {Halliwell}},\
  and\ \bibinfo {author} {\bibfnamefont {T.}~\bibnamefont {Venkataraman}},\
  }\bibfield  {title} {\bibinfo {title} {{Assessing the dynamical assumptions
  in Tsirelson inequality tests of non-classicality in harmonic oscillators}},\
  }\bibinfo {note} {arXiv:2509.03166}\BibitemShut {NoStop}%
\bibitem [{\citenamefont {Hofmann}(2017)}]{Hof17Quantum}%
  \BibitemOpen
  \bibfield  {author} {\bibinfo {author} {\bibfnamefont {H.~F.}\ \bibnamefont
  {Hofmann}},\ }\bibfield  {title} {\bibinfo {title} {{Quantum interference of
  position and momentum: A particle propagation paradox}},\ }\href
  {https://doi.org/10.1103/PhysRevA.96.020101} {\bibfield  {journal} {\bibinfo
  {journal} {Phys. Rev. A}\ }\textbf {\bibinfo {volume} {96}},\ \bibinfo
  {pages} {020101(R)} (\bibinfo {year} {2017})}\BibitemShut {NoStop}%
\bibitem [{\citenamefont {Hofmann}(2021)}]{Hof21How}%
  \BibitemOpen
  \bibfield  {author} {\bibinfo {author} {\bibfnamefont {H.~F.}\ \bibnamefont
  {Hofmann}},\ }\bibfield  {title} {\bibinfo {title} {{How to put quantum
  particles on magic bullet trajectories that can hit two targets without a
  clear line-of-sight}},\ }\href {https://doi.org/10.1038/s41598-021-87025-0}
  {\bibfield  {journal} {\bibinfo  {journal} {Sci. Rep.}\ }\textbf {\bibinfo
  {volume} {11}},\ \bibinfo {pages} {7964} (\bibinfo {year}
  {2021})}\BibitemShut {NoStop}%
\bibitem [{\citenamefont {Ono}\ \emph {et~al.}(2023)\citenamefont {Ono},
  \citenamefont {Samantarray},\ and\ \citenamefont {Rarity}}]{Ono2023}%
  \BibitemOpen
  \bibfield  {author} {\bibinfo {author} {\bibfnamefont {T.}~\bibnamefont
  {Ono}}, \bibinfo {author} {\bibfnamefont {N.}~\bibnamefont {Samantarray}},\
  and\ \bibinfo {author} {\bibfnamefont {J.~G.}\ \bibnamefont {Rarity}},\
  }\bibfield  {title} {\bibinfo {title} {{Controlling and measuring a
  superposition of position and momentum}},\ }\href
  {https://doi.org/10.1103/physreva.108.012215} {\bibfield  {journal} {\bibinfo
   {journal} {Phys. Rev. A}\ }\textbf {\bibinfo {volume} {108}},\ \bibinfo
  {pages} {012215} (\bibinfo {year} {2023})}\BibitemShut {NoStop}%
\bibitem [{\citenamefont {Trillo}\ \emph {et~al.}(2023)\citenamefont {Trillo},
  \citenamefont {Le},\ and\ \citenamefont {Navascués}}]{Trillo2023}%
  \BibitemOpen
  \bibfield  {author} {\bibinfo {author} {\bibfnamefont {D.}~\bibnamefont
  {Trillo}}, \bibinfo {author} {\bibfnamefont {T.~P.}\ \bibnamefont {Le}},\
  and\ \bibinfo {author} {\bibfnamefont {M.}~\bibnamefont {Navascués}},\
  }\bibfield  {title} {\bibinfo {title} {{Quantum advantages for transportation
  tasks - projectiles, rockets and quantum backflow}},\ }\href
  {https://doi.org/10.1038/s41534-023-00739-z} {\bibfield  {journal} {\bibinfo
  {journal} {npj Quant. Inf.}\ }\textbf {\bibinfo {volume} {9}},\ \bibinfo
  {pages} {69} (\bibinfo {year} {2023})}\BibitemShut {NoStop}%
\bibitem [{\citenamefont {Eliezer}\ \emph {et~al.}(2020)\citenamefont
  {Eliezer}, \citenamefont {Zacharias},\ and\ \citenamefont
  {Bahabad}}]{Eliezer2020}%
  \BibitemOpen
  \bibfield  {author} {\bibinfo {author} {\bibfnamefont {Y.}~\bibnamefont
  {Eliezer}}, \bibinfo {author} {\bibfnamefont {T.}~\bibnamefont {Zacharias}},\
  and\ \bibinfo {author} {\bibfnamefont {A.}~\bibnamefont {Bahabad}},\
  }\bibfield  {title} {\bibinfo {title} {{Observation of optical backflow}},\
  }\href {https://doi.org/10.1364/optica.371494} {\bibfield  {journal}
  {\bibinfo  {journal} {Optica}\ }\textbf {\bibinfo {volume} {7}},\ \bibinfo
  {pages} {72} (\bibinfo {year} {2020})}\BibitemShut {NoStop}%
\bibitem [{\citenamefont {Daniel}\ \emph {et~al.}(2022)\citenamefont {Daniel},
  \citenamefont {Ghosh}, \citenamefont {Gorzkowski},\ and\ \citenamefont
  {Lapkiewicz}}]{Daniel2022}%
  \BibitemOpen
  \bibfield  {author} {\bibinfo {author} {\bibfnamefont {A.}~\bibnamefont
  {Daniel}}, \bibinfo {author} {\bibfnamefont {B.}~\bibnamefont {Ghosh}},
  \bibinfo {author} {\bibfnamefont {B.}~\bibnamefont {Gorzkowski}},\ and\
  \bibinfo {author} {\bibfnamefont {R.}~\bibnamefont {Lapkiewicz}},\ }\bibfield
   {title} {\bibinfo {title} {{Demonstrating backflow in classical two beams'
  interference}},\ }\href {https://doi.org/10.1088/1367-2630/aca70b} {\bibfield
   {journal} {\bibinfo  {journal} {New J. Phys.}\ }\textbf {\bibinfo {volume}
  {24}},\ \bibinfo {pages} {123011} (\bibinfo {year} {2022})}\BibitemShut
  {NoStop}%
\bibitem [{\citenamefont {Ghosh}\ \emph {et~al.}(2023)\citenamefont {Ghosh},
  \citenamefont {Daniel}, \citenamefont {Gorzkowski},\ and\ \citenamefont
  {Lapkiewicz}}]{Ghosh2023}%
  \BibitemOpen
  \bibfield  {author} {\bibinfo {author} {\bibfnamefont {B.}~\bibnamefont
  {Ghosh}}, \bibinfo {author} {\bibfnamefont {A.}~\bibnamefont {Daniel}},
  \bibinfo {author} {\bibfnamefont {B.}~\bibnamefont {Gorzkowski}},\ and\
  \bibinfo {author} {\bibfnamefont {R.}~\bibnamefont {Lapkiewicz}},\ }\bibfield
   {title} {\bibinfo {title} {{Azimuthal backflow in light carrying orbital
  angular momentum}},\ }\href {https://doi.org/10.1364/optica.495710}
  {\bibfield  {journal} {\bibinfo  {journal} {Optica}\ }\textbf {\bibinfo
  {volume} {10}},\ \bibinfo {pages} {1217} (\bibinfo {year}
  {2023})}\BibitemShut {NoStop}%
\bibitem [{\citenamefont {Zhang}\ \emph {et~al.}(2025)\citenamefont {Zhang},
  \citenamefont {Huang}, \citenamefont {Dong}, \citenamefont {Rong},
  \citenamefont {Xu}, \citenamefont {Gu},\ and\ \citenamefont
  {Xiao}}]{ZHD+25Observation}%
  \BibitemOpen
  \bibfield  {author} {\bibinfo {author} {\bibfnamefont {Z.-F.}\ \bibnamefont
  {Zhang}}, \bibinfo {author} {\bibfnamefont {P.-F.}\ \bibnamefont {Huang}},
  \bibinfo {author} {\bibfnamefont {S.-C.}\ \bibnamefont {Dong}}, \bibinfo
  {author} {\bibfnamefont {Y.-X.}\ \bibnamefont {Rong}}, \bibinfo {author}
  {\bibfnamefont {J.-S.}\ \bibnamefont {Xu}}, \bibinfo {author} {\bibfnamefont
  {Y.-J.}\ \bibnamefont {Gu}},\ and\ \bibinfo {author} {\bibfnamefont
  {Y.}~\bibnamefont {Xiao}},\ }\bibfield  {title} {\bibinfo {title}
  {Observation of single-photon azimuthal backflow with weak measurement},\
  }\href {https://doi.org/10.1364/OL.540905} {\bibfield  {journal} {\bibinfo
  {journal} {Opt. Lett.}\ }\textbf {\bibinfo {volume} {50}},\ \bibinfo {pages}
  {333} (\bibinfo {year} {2025})}\BibitemShut {NoStop}%
\bibitem [{\citenamefont {Eveson}\ \emph {et~al.}(2005)\citenamefont {Eveson},
  \citenamefont {Fewster},\ and\ \citenamefont {Verch}}]{EFV05Quantum}%
  \BibitemOpen
  \bibfield  {author} {\bibinfo {author} {\bibfnamefont {S.~P.}\ \bibnamefont
  {Eveson}}, \bibinfo {author} {\bibfnamefont {C.~J.}\ \bibnamefont
  {Fewster}},\ and\ \bibinfo {author} {\bibfnamefont {R.}~\bibnamefont
  {Verch}},\ }\bibfield  {title} {\bibinfo {title} {{Quantum Inequalities in
  Quantum Mechanics}},\ }\href {https://doi.org/10.1007/s00023-005-0197-9}
  {\bibfield  {journal} {\bibinfo  {journal} {Ann. Henri Poincar\'e}\ }\textbf
  {\bibinfo {volume} {6}},\ \bibinfo {pages} {1} (\bibinfo {year}
  {2005})}\BibitemShut {NoStop}%
\bibitem [{\citenamefont {Penz}\ \emph {et~al.}(2006)\citenamefont {Penz},
  \citenamefont {Gr{\"{u}}bl}, \citenamefont {Kreidl},\ and\ \citenamefont
  {Wagner}}]{PGKW06new}%
  \BibitemOpen
  \bibfield  {author} {\bibinfo {author} {\bibfnamefont {M.}~\bibnamefont
  {Penz}}, \bibinfo {author} {\bibfnamefont {G.}~\bibnamefont {Gr{\"{u}}bl}},
  \bibinfo {author} {\bibfnamefont {S.}~\bibnamefont {Kreidl}},\ and\ \bibinfo
  {author} {\bibfnamefont {P.}~\bibnamefont {Wagner}},\ }\bibfield  {title}
  {\bibinfo {title} {{A new approach to quantum backflow}},\ }\href
  {https://doi.org/10.1088/0305-4470/39/2/012} {\bibfield  {journal} {\bibinfo
  {journal} {J. Phys. A: Math. Gen.}\ }\textbf {\bibinfo {volume} {39}},\
  \bibinfo {pages} {423} (\bibinfo {year} {2006})}\BibitemShut {NoStop}%
\bibitem [{\citenamefont {Fewster}\ and\ \citenamefont
  {{Kirk-Karakaya}}(2025)}]{FK--Repeated}%
  \BibitemOpen
  \bibfield  {author} {\bibinfo {author} {\bibfnamefont {C.~J.}\ \bibnamefont
  {Fewster}}\ and\ \bibinfo {author} {\bibfnamefont {H.~J.}\ \bibnamefont
  {{Kirk-Karakaya}}},\ }\bibfield  {title} {\bibinfo {title} {Repeated quantum
  backflow and overflow},\ }\href {https://doi.org/10.1098/rspa.2025.0577}
  {\bibfield  {journal} {\bibinfo  {journal} {Proc. R. Soc. A}\ }\textbf
  {\bibinfo {volume} {481}},\ \bibinfo {pages} {20250577} (\bibinfo {year}
  {2025})}\BibitemShut {NoStop}%
\bibitem [{\citenamefont {Yearsley}\ \emph {et~al.}(2012)\citenamefont
  {Yearsley}, \citenamefont {Halliwell}, \citenamefont {Hartshorn},\ and\
  \citenamefont {Whitby}}]{Yearsley2012}%
  \BibitemOpen
  \bibfield  {author} {\bibinfo {author} {\bibfnamefont {J.~M.}\ \bibnamefont
  {Yearsley}}, \bibinfo {author} {\bibfnamefont {J.~J.}\ \bibnamefont
  {Halliwell}}, \bibinfo {author} {\bibfnamefont {R.}~\bibnamefont
  {Hartshorn}},\ and\ \bibinfo {author} {\bibfnamefont {A.}~\bibnamefont
  {Whitby}},\ }\bibfield  {title} {\bibinfo {title} {{Analytical examples,
  measurement models, and classical limit of quantum backflow}},\ }\href
  {https://doi.org/10.1103/physreva.86.042116} {\bibfield  {journal} {\bibinfo
  {journal} {Phys. Rev. A}\ }\textbf {\bibinfo {volume} {86}},\ \bibinfo
  {pages} {042116} (\bibinfo {year} {2012})}\BibitemShut {NoStop}%
\bibitem [{\citenamefont {Halliwell}\ \emph {et~al.}(2013)\citenamefont
  {Halliwell}, \citenamefont {Gillman}, \citenamefont {Lennon}, \citenamefont
  {Patel},\ and\ \citenamefont {Ramirez}}]{HGL+13Quantum}%
  \BibitemOpen
  \bibfield  {author} {\bibinfo {author} {\bibfnamefont {J.~J.}\ \bibnamefont
  {Halliwell}}, \bibinfo {author} {\bibfnamefont {E.}~\bibnamefont {Gillman}},
  \bibinfo {author} {\bibfnamefont {O.}~\bibnamefont {Lennon}}, \bibinfo
  {author} {\bibfnamefont {M.}~\bibnamefont {Patel}},\ and\ \bibinfo {author}
  {\bibfnamefont {I.}~\bibnamefont {Ramirez}},\ }\bibfield  {title} {\bibinfo
  {title} {{Quantum backflow states from eigenstates of the regularized current
  operator}},\ }\href {https://doi.org/10.1088/1751-8113/46/47/475303}
  {\bibfield  {journal} {\bibinfo  {journal} {J. Phys. A: Math. Theor.}\
  }\textbf {\bibinfo {volume} {46}},\ \bibinfo {pages} {475303} (\bibinfo
  {year} {2013})}\BibitemShut {NoStop}%
\bibitem [{\citenamefont {Palmero}\ \emph {et~al.}(2013)\citenamefont
  {Palmero}, \citenamefont {Torrontegui}, \citenamefont {Muga},\ and\
  \citenamefont {Modugno}}]{Palmero2013}%
  \BibitemOpen
  \bibfield  {author} {\bibinfo {author} {\bibfnamefont {M.}~\bibnamefont
  {Palmero}}, \bibinfo {author} {\bibfnamefont {E.}~\bibnamefont
  {Torrontegui}}, \bibinfo {author} {\bibfnamefont {J.~G.}\ \bibnamefont
  {Muga}},\ and\ \bibinfo {author} {\bibfnamefont {M.}~\bibnamefont
  {Modugno}},\ }\bibfield  {title} {\bibinfo {title} {{Detecting quantum
  backflow by the density of a Bose-Einstein condensate}},\ }\href
  {https://doi.org/10.1103/physreva.87.053618} {\bibfield  {journal} {\bibinfo
  {journal} {Phys. Rev. A}\ }\textbf {\bibinfo {volume} {87}},\ \bibinfo
  {pages} {053618} (\bibinfo {year} {2013})}\BibitemShut {NoStop}%
\bibitem [{\citenamefont {Miller}\ \emph {et~al.}(2021)\citenamefont {Miller},
  \citenamefont {Woo}, \citenamefont {Dumke},\ and\ \citenamefont
  {Paterek}}]{Miller2021}%
  \BibitemOpen
  \bibfield  {author} {\bibinfo {author} {\bibfnamefont {M.}~\bibnamefont
  {Miller}}, \bibinfo {author} {\bibfnamefont {C.~Y.}\ \bibnamefont {Woo}},
  \bibinfo {author} {\bibfnamefont {R.}~\bibnamefont {Dumke}},\ and\ \bibinfo
  {author} {\bibfnamefont {T.}~\bibnamefont {Paterek}},\ }\bibfield  {title}
  {\bibinfo {title} {{Experiment-friendly formulation of quantum backflow}},\
  }\href {https://doi.org/10.22331/q-2021-01-11-379} {\bibfield  {journal}
  {\bibinfo  {journal} {Quantum}\ }\textbf {\bibinfo {volume} {5}},\ \bibinfo
  {pages} {379} (\bibinfo {year} {2021})}\BibitemShut {NoStop}%
\bibitem [{\citenamefont {Barbier}\ and\ \citenamefont
  {Goussev}(2021)}]{BG21experiment}%
  \BibitemOpen
  \bibfield  {author} {\bibinfo {author} {\bibfnamefont {M.}~\bibnamefont
  {Barbier}}\ and\ \bibinfo {author} {\bibfnamefont {A.}~\bibnamefont
  {Goussev}},\ }\bibfield  {title} {\bibinfo {title} {{On the
  experiment-friendly formulation of quantum backflow}},\ }\href
  {https://doi.org/10.22331/q-2021-09-07-536} {\bibfield  {journal} {\bibinfo
  {journal} {Quantum}\ }\textbf {\bibinfo {volume} {5}},\ \bibinfo {pages}
  {536} (\bibinfo {year} {2021})}\BibitemShut {NoStop}%
\bibitem [{\citenamefont {Chremmos}(2024)}]{Chr24Design}%
  \BibitemOpen
  \bibfield  {author} {\bibinfo {author} {\bibfnamefont {I.}~\bibnamefont
  {Chremmos}},\ }\bibfield  {title} {\bibinfo {title} {{Design of quantum
  backflow in the complex plane}},\ }\href
  {https://doi.org/10.1088/1751-8121/ad1aca} {\bibfield  {journal} {\bibinfo
  {journal} {J. Phys. A: Math. Theor.}\ }\textbf {\bibinfo {volume} {57}},\
  \bibinfo {pages} {055301} (\bibinfo {year} {2024})}\BibitemShut {NoStop}%
\bibitem [{\citenamefont {Goussev}\ and\ \citenamefont
  {Morozov}(2025)}]{Goussev2025}%
  \BibitemOpen
  \bibfield  {author} {\bibinfo {author} {\bibfnamefont {A.}~\bibnamefont
  {Goussev}}\ and\ \bibinfo {author} {\bibfnamefont {G.~V.}\ \bibnamefont
  {Morozov}},\ }\bibfield  {title} {\bibinfo {title} {{Searching for Bloch wave
  packets with almost definite momentum direction}},\ }\href
  {https://doi.org/10.1103/k9km-k3dm} {\bibfield  {journal} {\bibinfo
  {journal} {Phys. Rev. A}\ }\textbf {\bibinfo {volume} {112}},\ \bibinfo
  {pages} {032223} (\bibinfo {year} {2025})}\BibitemShut {NoStop}%
\bibitem [{\citenamefont {Goussev}(2020)}]{Gou20Probability}%
  \BibitemOpen
  \bibfield  {author} {\bibinfo {author} {\bibfnamefont {A.}~\bibnamefont
  {Goussev}},\ }\bibfield  {title} {\bibinfo {title} {{Probability backflow for
  correlated quantum states}},\ }\href
  {https://doi.org/10.1103/PhysRevResearch.2.033206} {\bibfield  {journal}
  {\bibinfo  {journal} {Phys. Rev. Research}\ }\textbf {\bibinfo {volume}
  {2}},\ \bibinfo {pages} {033206} (\bibinfo {year} {2020})}\BibitemShut
  {NoStop}%
\bibitem [{\citenamefont {Bevington}\ and\ \citenamefont
  {Robinson}(2003)}]{Bevington2003}%
  \BibitemOpen
  \bibfield  {author} {\bibinfo {author} {\bibfnamefont {P.}~\bibnamefont
  {Bevington}}\ and\ \bibinfo {author} {\bibfnamefont {D.}~\bibnamefont
  {Robinson}},\ }\href@noop {} {\emph {\bibinfo {title} {{Data Reduction and
  Error Analysis for the Physical Sciences}}}}\ (\bibinfo  {publisher}
  {McGraw-Hill Education},\ \bibinfo {year} {2003})\BibitemShut {NoStop}%
\bibitem [{\citenamefont {Muga}\ and\ \citenamefont
  {Leavens}(2000)}]{ML00Arrival}%
  \BibitemOpen
  \bibfield  {author} {\bibinfo {author} {\bibfnamefont {J.~G.}\ \bibnamefont
  {Muga}}\ and\ \bibinfo {author} {\bibfnamefont {C.~R.}\ \bibnamefont
  {Leavens}},\ }\bibfield  {title} {\bibinfo {title} {{Arrival time in quantum
  mechanics}},\ }\href {https://doi.org/10.1016/S0370-1573(00)00047-8}
  {\bibfield  {journal} {\bibinfo  {journal} {Phys. Rep.}\ }\textbf {\bibinfo
  {volume} {338}},\ \bibinfo {pages} {353} (\bibinfo {year}
  {2000})}\BibitemShut {NoStop}%
\end{thebibliography}
\end{document}